\documentclass[fleqn,10pt,twocolumn]{wlscirep}
\usepackage[utf8]{inputenc}
\usepackage[T1]{fontenc}

\usepackage{epstopdf}
\usepackage{subfigure}
\usepackage{stfloats}
\usepackage{cite}
\usepackage{amsmath,amssymb,amsfonts}
\usepackage{graphicx}
\usepackage{subcaption}
\usepackage{textcomp}
\usepackage{xcolor}
\usepackage{flushend}
\usepackage{algorithm}
\usepackage{algpseudocode}
\usepackage{amsthm}

\newtheorem{remark}{Remark}

\DeclareMathAlphabet{\mathcal}{OMS}{cmsy}{m}{n}

\title{Time-Modulated Intelligent Reflecting Surfaces for Integrated Sensing, Communication and Security: A Generative AI Design Framework}

\author{Zhihao Tao}
\author{Athina Petropulu}
\author{H. Vincent Poor$^*$}
\affil{Electrical and Computer Engineering, Rutgers University \\
$^*$Electrical and Computer Engineering, Princeton University}

\begin{abstract}
We propose a novel approach to achieve physical layer security for integrated sensing and communication (ISAC) systems operating in the presence of targets that may be eavesdroppers. The system is aided  by a time-modulated intelligent reflecting surface (TM-IRS), which is 
configured to preserve the  integrity of the transmitted data at one or more legitimate communication users (CUs) while making them appear scrambled in  all other directions. 
The TM-IRS design leverages a generative flow network (GFlowNet) framework to learn a stochastic policy that samples high-performing TM-IRS configurations from a vast discrete parameter space. Specifically, we begin by formulating the achievable sum rate for the legitimate CUs and the beampattern gain toward the target direction, based on which we construct reward functions for GFlowNets that jointly capture both communication and sensing performance. The TM-IRS design is modeled as a deterministic Markov decision process (MDP), where each terminal state corresponds to a complete configuration of TM-IRS parameters. GFlowNets, parametrized by deep neural networks are employed to learn a stochastic policy that samples TM-IRS parameter sets with probability proportional to their associated reward.
Experimental results demonstrate the effectiveness of the proposed GFlowNet-based method in integrating sensing, communication and security simultaneously, and also exhibit significant sampling efficiency as compared to the exhaustive combinatorial search and enhanced robustness against the existing benchmarks of physical layer security.

\vspace{1em}
\noindent\textbf{Keywords:} Dual-Function Radar-Communication (DFRC), intelligent reflecting surface (IRS), time modulation, physical layer security (PLS), generative AI (GenAI), GFlowNets.
\end{abstract}

\begin{document}

\flushbottom
\maketitle
%
%
\thispagestyle{empty}

\section{Introduction}
The explosive growth in the number of wireless devices and the increasing demand for both high data rates and ubiquitous environmental awareness have propelled integrated sensing and communication (ISAC) to the forefront of 6G research and standardization. ISAC systems aim to jointly perform communication and sensing tasks using the same waveform, hardware, or spectrum, thereby reducing cost, improving spectral efficiency, and enabling tighter coordination between devices and their environments~\cite{liu2020joint, mishra2019toward, zhang2021overview}. By fusing these traditionally separate functionalities, ISAC paves the way for transformative applications such as autonomous driving, smart cities/factories, human-device interaction, and surveillance systems~\cite{sun2020mimo, wymeersch2017mmwave}. 

Within the broader ISAC paradigm, dual-function radar-communication (DFRC) systems have emerged as a compelling architecture that uses a fully-shared hardware platform and a unified transmit waveform to simultaneously probe the physical environment and convey data to communication users (CUs). DFRC designs benefit from streamlined hardware, coherent integration, and real-time synchronization between radar and communication operations~\cite{hassanien2019dual, wang2025dfrcotfs, wang2025ISACMIMO}. Orthogonal frequency-division multiplexing (OFDM)-based DFRC systems \cite{xu2023bandwidth}, in particular, offer high flexibility, wide bandwidth, and compatibility with existing multi-carrier communication standards, making them a natural candidate for ISAC implementations. 
Despite these advantages, DFRC systems are increasingly recognized to suffer from critical security vulnerabilities at the physical layer. Since the same signal is used for both radar and communication purposes, radar targets may inadvertently or maliciously intercept communication data.  Consequently, conventional DFRC designs are vulnerable to eavesdropping attacks by the targets~\cite{xu2023secure, su2021secure, su2023sensing, hua2023secure}. Therefore, developing physical layer security (PLS) mechanisms that can safeguard communication while enabling effective target sensing is crucial.

PLS exploits the physical characteristics of the wireless medium, such as channel fading, noise, interference, and spatial diversity, to complement, or in some circumstances, replace higher-layer cryptographic techniques~\cite{Shannon1949Comm,Wyner1975Wire,poor2017wireless, dong2009improving}. Among the many PLS mechanisms proposed for securing DFRC systems, directional modulation (DM) has attracted particular interest because it embeds information directly in the spatial signature of the transmitted waveform: a receiver aligned with the intended steering direction observes an undistorted constellation, whereas other directions see a scrambled one~\cite{daly2009dire,qiu2023decomposed,tao2024tma}. Compared with other PLS approaches such as secrecy rate maximization \cite{lv2015secrecy, gong2016millimeter} or artificial-noise injection \cite{zhang2019AN,wang2017AN}, DM can offer comparable secrecy in a more cost-effective and energy-efficient manner \cite{su2022secure}.

DM implementations have been proposed for fully digital or hybrid beamforming architectures with multiple radio-frequency (RF) chains and fine-grained phase control at each antenna element or each transmitted symbol \cite{li2019performance, Ottersten2016, Alodeh2016DM, su2021secure, su2022secure}. Also,
many methods \cite{li2023irs, su2022secure, Spilios2024ISAC} necessitate full channel state information (CSI) on  the eavesdropper as well as the legitimate users.
There are DM approaches that   exploit high-speed phase manipulation to control the detectability of the transmitted signal in different directions. For example, \cite{zhao2024low} employs rapid sidelobe time modulation using 1-bit phase-shifter toggling, which redistributes sidelobe energy over time and thereby reduces detectability in unintended directions while preserving main-lobe communication performance. \cite{paidimarri2020spatio} employs microsecond-scale fast beam switching to achieve precise spatio-temporal beam control, significantly reducing sidelobes and thereby increasing interception difficulty. \cite{tollefson2015out} adopts symbol-level phase modulation across antenna elements to intentionally distort received constellations in undesired directions, achieving directional physical-layer encryption. Although these schemes also offer promising PLS mechanisms, they are designed exclusively for single-carrier waveforms, and their extension to multi-carrier OFDM-based DFRC remains nontrivial.

DM enabled by a
time-modulated array (TMA)  driven by OFDM signals has also been considered as a    hardware cost-efficient solution for securing DFRC systems, while also enabling  high data rates through OFDM \cite{xu2023secure}, without the need for multiple RF chains or CSI knowledge~\cite{tvt2019time, tao2024tma, tao2025twc}.
By using single-pole-single-throw (SPST), the TMA  periodically connects   and disconnects antennas to one single RF chain, generating controllable harmonics whose periods are aligned with the OFDM symbol duration. The harmonics cause the symbols initially placed on each subcarrier to spread across all subcarriers.
As a result, each subcarrier of the transmitted OFDM signal carries a weighted mixture of all original symbols, with mixing coefficients determined by the TMA parameters—namely, the connection times or states, and the corresponding state durations. This mixing effectively scrambles the transmitted symbols. In the absence of noise, the scrambling towards an intended direction can be eliminated by a rule-based design \cite{tvt2019time}, and this can be implemented with low complexity. 

In this work, our focus is on OFDM-based  DFRC \cite{xu2023bandwidth}, which can achieve high communication rates. For such systems,  TMA is a natural fit to implement PLS.
Despite the  advantages of the TMA DM approach,  the periodic deactivation of antenna elements degrades the systems  energy utilization efficiency
~\cite{hou2023energy}. To address this issue, recent research \cite{xu2025tmirs} shifts time modulation to an intelligent reflecting surface (IRS). IRS is a passive metasurface composed of programmable elements that dynamically adjust the phase of incident electromagnetic waves to realize beamforming \cite{wu2019towards, wu2019intelligent}. By exerting the periodic TM on each IRS element, the system in \cite{xu2025tmirs} is designed to implement a 3D directional modulation. The large aperture of an IRS delivers substantial beamforming gain that compensates for power loss of TMA during element deactivation. In \cite{xu2025tmirs}, the TM-IRS parameters are determined using simple, closed-form rules that are straightforward to implement. However, this approach does not account for noise and guarantees undistorted signal reception only in a single CU direction. Extending this approach to support communication with multiple users is challenging—a significant limitation given that multi-user scenarios are common in modern wireless systems, particularly in ISAC settings.

In this paper, we formulate a TM-IRS–assisted DFRC system driven by OFDM waveforms and propose a TM-IRS design approach that is not rule-based and can accommodate noise as well as multiple CUs. We first assume that 
the target/eavesdropper's location lies within a region of the 3D space. During the target tracking stage, this region is determined based on previous detections and predicted target positions.
%
We define the secrecy rate as the difference between the CU sum rate and the  eavesdropper sum rate,
and  design the TM  and IRS parameters to maximize the minimal (worst-case) secrecy rate across all possible locations within the suspected target region subject to sensing constraints.
The main novelty of our work is a new  generative AI (GenAI)-based framework for solving the design problem. In particular, 
we adopt a sampling-based strategy that selects high-quality TM-IRS parameter configurations from a discretized space of all possible IRS element on/off and phase settings. The quality of each configuration is evaluated through a designed reward function that contains the above defined secrecy rate and the desired  beampattern gain. We first formulate the TM-IRS design task as a deterministic Markov decision process (MDP), in which each terminal state corresponds to a complete TM-IRS configuration over all IRS elements. To solve this problem, we  employ generative flow networks (GFlowNets)~\cite{bengio2021flow, bengio2023gflownet, zhang2022generative, tao2025secure, tao2025meta}, a class of unsupervised generative models that utilize the flow matching principle to learn stochastic policies for sampling structured configurations with probability proportional to the predefined reward. A deep neural network-based GFlowNet is trained offline to model the trajectory flow in the MDP and enforce flow matching. Once trained, the model can be deployed online to sample TM-IRS configurations that maximize the communication sum rate while ensuring that the security and the radar sensing performance are satisfied. Note that the computational complexity of the proposed GFlowNet is dominated by the training process while the deployment stage involves significantly lower computational overhead.

Experimental results validate the effectiveness of the proposed approach, showing that the learned GFlowNet-based policy generates TM-IRS patterns that support multiple users, and achieve more robust security performance against existing PLS benchmarks. Moreover, the sampling policy is stochastic and remains hidden from adversaries, significantly increasing the difficulty of interception or reverse-engineering. Notably, the GFlowNet achieves strong performance even when trained on fewer than $0.000001\%$ of all possible configurations, highlighting its sampling efficiency compared to exhaustive combinatorial search.





The remainder of the paper is organized as follows. Section~\ref{system_model} describes the system model, including the TM-IRS-assisted DFRC architecture, OFDM transmission, and the performance metrics used for evaluating communication and sensing. Section \ref{problem} gives the problem formulation in a practical scenario. Section~\ref{gflownet} presents the proposed GFlowNet-based TM-IRS design framework, detailing the MDP formulation, reward construction, and algorithm procedure. Section~\ref{experiments} provides simulation results that compare the proposed method with baseline approaches under various DFRC settings. Finally, Section~\ref{conclusion} concludes the paper and outlines potential directions for future research.

\textit{Notation:} Throughout the paper, bold uppercase letters (e.g., $\mathbf{X}$), bold lowercase letters (e.g., $\mathbf{x}$), and lowercase letters (e.g., \(x\)) represent matrices, column vectors, and scalars, respectively. Superscripts $(\cdot)^T$, $(\cdot)^*$, and $(\cdot)^\dagger$ denote the transpose, complex conjugate, and Hermitian transpose, respectively. $\otimes$ denotes the Kronecker product. The notation $\text{tr}(\mathbf{X})$, $|\mathbf{X}|$, and $\|\mathbf{X}\|$ indicate the trace, modulus, and $\ell_2$-norm of $\mathbf{X}$, respectively. The expectation operator is denoted by $\mathbb{E}[\cdot]$.

To improve readability, we summarize in Table~\ref{table:notation} the key parameters and symbols used throughout the paper, including mainly system dimensions and TM-IRS configuration variables.

\begin{table}[t]
\centering
\caption{Summary of Main Symbols and Definitions}
\label{table:notation}
\renewcommand{\arraystretch}{1.1}
\setlength{\tabcolsep}{3.5pt}
\begin{tabular}{l p{5.8cm}}
\hline
\textbf{Symbol} & \textbf{Meaning} \\
\hline
$M_x, M_z$ & Number of IRS elements along $x$- and $z$-axes \\
$K$ & Number of OFDM subcarriers \\
$N_t$ & Number of BS ULA antennas \\
$N_s$ & Number of samples within one OFDM symbol duration \\
$N_p$ & Number of discretized angle pairs in $\Psi$\\
$\Psi$ & Suspected target region (sensing region) \\
$\theta, \phi$ & IRS elevation and azimuth angles \\
$\xi$ & BS ULA azimuth angle \\
$f_c, f_s, T_s$ & OFDM carrier frequency, subcarrier spacing and OFDM symbol duration \\
\hline
$c_{mn}$ & Phase shift of IRS element $(m,n)$ \\
$\tau_{mn}^o$ & Turn-on instant of IRS element $(m,n)$ within period $T_s$ \\
$\Delta \tau_{mn}$ & Activation duration of IRS element $(m,n)$ \\
$Q_1, Q_2, Q_3$ & Number of quantization levels for $c_{mn}$, $\tau^o_{mn}$, and $\Delta\tau_{mn}$ \\
$V_j$ & $j$-th harmonic coefficient produced by TM-IRS \\
\hline
$\overline{\gamma_r}$ & Average radar beampattern gain induced by TM-IRS \\
$\gamma_{\text{th}}$ & Minimum required radar-sensing threshold \\
$\rho_u$ & Allowed phase deviation for constellation recovery at CU \\
$\mathcal{R}(\mathbf{s})$ & Terminal-state reward \\
$\ln Z$ & log-partition estimate \\
\hline
\end{tabular}
\end{table}


\section{System Model}\label{system_model}
\begin{figure}[t]
\centerline{\includegraphics[width=3.0in]{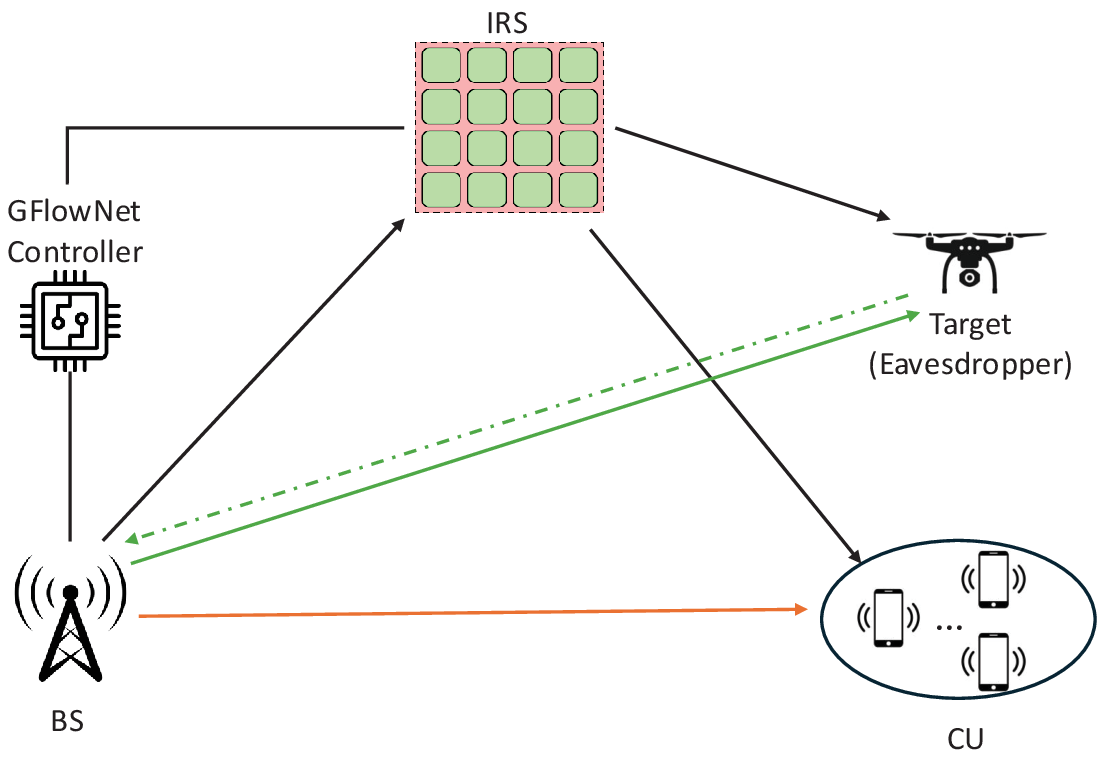}}
\caption{Illustration of the TM-IRS-assisted DFRC system, where the IRS is placed adjacent to the BS to allow collaborative beamforming and a GFlowNet controller is adopted to configure the IRS.
}
\label{fig1}
\end{figure}

The considered DFRC system is illustrated in Fig.~\ref{fig1}. It consists of a base station (BS) equipped with a uniform linear array (ULA) that transmits OFDM signals to both legitimate communication users (CUs) and a radar target, which is also considered a potential eavesdropper.
Both the CUs and the eavesdropper are in the line-of-sight (LOS) of the BS.
During transmission, the BS employs beamforming to direct the signal toward the IRS. Upon reflection from the IRS, the signal reaches both the legitimate CUs and the eavesdropper. 
%
For communication, the CUs and the eavesdropper can receive the transmitted signals from both the BS and IRS side.
{For radar sensing, the BS primarily relies on echoes received through the LOS path to estimate the target parameters\footnote{{Even with only the LOS return, the BS can reliably estimate the target parameters through classical array processing methods such as matched filtering or beam scanning}}, as the non-line-of-sight (NLOS) echoes—arriving after reflection from the IRS—are significantly attenuated due to the double fading effect~\cite{sharma2023performance}.}

%
The IRS consists of $M_x \times M_z$ reflecting elements. Let $(\theta_b, \phi_b)$ denote the elevation and azimuth angles of the ULA transmitter from the IRS's perspective. {Due to the sub-wavelength size of each IRS element and the overall compactness of the surface~\cite{hua2023secure}, the IRS is modeled as a point target from the BS perspective, and its direction is denoted by $\xi_{\text{IRS}}$. 
To simplify notation, we initially consider a single legitimate CU and denote its direction relative to the IRS and the BS as $(\theta_u, \phi_u)$ and $\xi_u$, respectively.
We assume that $\xi_{\text{IRS}}$, $\xi_u$ are known to the BS, and both $(\theta_b, \phi_b)$ and $(\theta_u, \phi_u)$ are known to the IRS since DM necessitates the location knowledge of the legitimate CU. All elements of the ULA and IRS are spaced at half the carrier wavelength, i.e., $\lambda/2$. The phase shifter and SPST switch applied to each element of IRS are controlled by the proposed GFlowNet in this paper. 
{The channels of each CU is assumed to be known by the CU receiver, so that they  can be compensated for.} We also assume that the eavesdropper receiver knows its channel to the IRS and the BS, and can perfectly compensate for it, such that channel effects do not contribute to additional signal scrambling. In this sense, we consider the most challenging scenario—attempting to confuse an eavesdropper who has extensive knowledge of the DFRC system.
}




{
As is common in DFRC system design~\cite{xu2024reconfigurable}, we assume that the system operates in both searching and tracking modes. In the searching mode, which is periodically invoked, the system performs coarse target estimation. In the tracking mode, it uses these estimates to carry out joint communication and sensing. Specifically, as long as the target remains within the mainlobe of the designed beampattern, it is continuously illuminated, enabling progressive refinement of target parameters. 
The proposed system primarily focuses on the tracking phase, assuming that an approximate target position region relative to the IRS is already available.}


\begin{remark}
In the considered system of Fig.~\ref{fig1}, we also observe that the received signal at the eavesdropper includes  a non-scrambled component originating from the LOS path. This path preserves the original data symbol and therefore may pose a potential security risk if its power is significant. However, the BS ULA adopted here steers its main-lobe toward the IRS, which suppresses the LOS component in other directions. As a result, the LOS component towards the eavesdropper remains small and the resulting leakage is non-dominant. Moreover, the optimized TM-IRS parameters can enhance NLOS components, further reducing the relative impact of the LOS path. In extreme scenarios where the LOS component is strong, a more sophisticated BS beamformer or precoder could be employed to further suppress LOS leakage, which is an interesting direction for future work. In this paper, our focus remains on the TM-IRS design rather than BS beamforming optimization.
\end{remark}

For the above described system,  the communication and sensing signal models are presented in the following subsections.

\subsection{Communication Model}
In the ULA, each antenna element is fed with an OFDM signal, which is expressed as
\begin{equation}\label{eq1}
    e(t) = \frac{1}{\sqrt{K}} \sum_{k=0}^{K-1} d(k) e^{j2\pi(f_c + k f_s)t}, \quad 0 \leq t < T_s,
\end{equation}
where $K$ is the number of subcarriers, $d(k)$ is the digitally modulated data symbol on the $k$-th subcarrier, which has been normalized to be zero-mean and unit-variance, $f_c$ is the carrier frequency, $f_s$ is the subcarrier spacing, and $T_s$ is the OFDM symbol duration. 
{On using a conjugate beamformer, i.e., setting the antenna weights to $w_n = e^{-j n \pi \cos (\xi_{\text{IRS}})}$,  the ULA beam is focused towards the IRS. This beamforming operation is implemented through a conventional analog phased-array architecture, where each antenna element applies a fixed phase shift to form the desired steering vector.} The radiated waveform toward angle $\xi$ with respect to the BS  equals
\begin{equation}
    r(t,\xi) = \frac{1}{\sqrt{N_t}} \sum_{n=0}^{N_t-1} e(t) w_n e^{j n \pi \cos (\xi)},
\end{equation}
where $N_t$ is the number of transmit antennas.

\begin{figure}[t]
    \centering

    \begin{minipage}{0.225\textwidth}
        \centering
        \includegraphics[width=\textwidth]{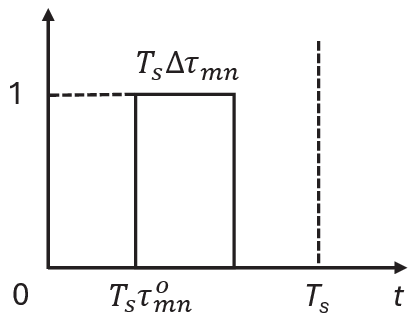}
        \vspace{0.5em}  
        \small (a)
    \end{minipage}
    \hfill
    \begin{minipage}{0.225\textwidth}
        \centering
        \includegraphics[width=\textwidth]{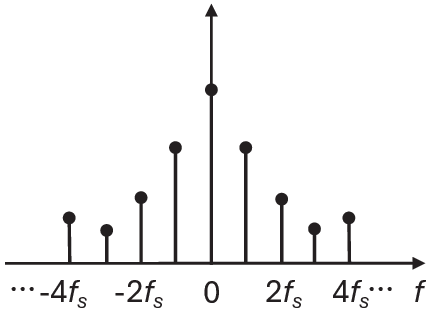}
        \vspace{0.5em}
        \small (b)
    \end{minipage}

    \caption{Illustrations of one period of the switch controlling function $U_{mn}(t)$: (a) time domain; (b) frequency domain.}
    
    \label{fig2}
\end{figure}

Each IRS unit is connected to a high-speed SPST switch and a phase shifter. The switches operate in two states: ``on'' and ``off.'' Let $U_{mn}(t)$ denote the periodic on/off switching function of the $(m,n)$-th IRS unit ($m=0,1,\cdots,M_x-1, n=0,1,\cdots,M_z-1$), with a period equal to $T_s$, as shown in Fig. \ref{fig2} (a). Also, let the normalized turn-on instant be $\tau^o_{mn} \in [0,1)$ and the normalized on-duration  $\Delta\tau_{mn} \in [0,1)$. The switching function $U_{mn}(t)$ is set to 1 when $t \in [T_s \tau^o_{mn}, T_s (\tau^o_{mn} + \Delta\tau_{mn})]$ and $0$ otherwise. This periodic square waveform can be expanded using its Fourier series as
\begin{equation}\label{eq2}
\begin{split}
    U_{mn}(t) = \sum_{l=-\infty}^{\infty} e^{j2\pi l f_s t} &\Delta\tau_{mn} \mathrm{sinc}(l\pi \Delta\tau_{mn}) \\
    &\times e^{-j l\pi (2\tau^o_{mn} + \Delta\tau_{mn})},
\end{split}
\end{equation}
where the harmonics introduced by time modulation are centered at integer multiples of $f_s$. The   magnitude of the harmonic components is shown in Fig. \ref{fig2} (b). 
Considering the receiver noise $z(t)$, {the radiated signal towards directions \((\theta, \phi)\) with respect to the IRS and direction $\xi$ with respect to the BS is
\begin{equation}
\begin{split}
    {y}(t, \theta, \phi, \xi) 
    &= \zeta_{\text{NLOS}} \mathbf{a}^T(\theta, \phi) \boldsymbol{\Phi} \mathbf{U}(t) \mathbf{a}(\theta_b, \phi_b) r(t,\xi_{\text{IRS}})
    \\
    &+ \zeta_{\text{LOS}} r(t,\xi) + z(t)
    \\
    &= \zeta_{\text{NLOS}} \mathbf{a}^T(\theta, \phi) \boldsymbol{\Phi} \mathbf{U}(t) \mathbf{a}(\theta_b, \phi_b) \sqrt{N_t} e(t)
    \\
    &+ \zeta_{\text{LOS}} \Gamma (\xi) e(t) + z(t),
\end{split}
\label{eq3}
\end{equation}
where $\zeta_{\text{NLOS}}$, $\zeta_{\text{LOS}}$ represent the NLOS and LOS path loss, respectively}; $\Gamma (\xi) = \frac{1}{\sqrt{N_t}} \sum_{n=0}^{N_t-1} w_n e^{j n \pi \cos (\xi)}$ represents the BS steering factor along the LOS direction $\xi$; $\mathbf{a}(\theta, \phi)$ is the IRS steering vector, given by
\begin{equation}
\begin{split}
    \mathbf{a}^T(\theta, \phi) &= [1, e^{-j\pi \sin \theta \cos \phi}, \ldots, e^{-j\pi (M_x-1) \sin \theta \cos \phi}] \\
    &\quad \otimes [1, e^{-j\pi \sin \theta \sin \phi}, \ldots, e^{-j\pi (M_z-1) \sin \theta \sin \phi}].
\end{split}
\end{equation}
The matrices $\boldsymbol{\Phi}$ and $\mathbf{U}(t)$ are diagonal, with each diagonal element corresponding to the unit-modulus phase shift $c_{mn}$ and the time modulation function $U_{mn}(t)$, respectively, of the $(m,n)$-th IRS element. 
%
%



Substituting \eqref{eq1} and \eqref{eq2} into \eqref{eq3} and reorganizing the terms yields
\begin{equation}
\begin{aligned}
    &y(t, \theta, \phi, \xi) = \sum_{k=0}^{K-1} d(k) e^{j2\pi(f_c + k f_s)t} \\
    &\times \left[{\frac{\zeta_{\text{NLOS}} \sqrt N_t}{\sqrt K}}  \sum_{l=-\infty}^{\infty} e^{j2\pi l f_s t} V(l, \Omega, \theta, \phi) + \frac{\zeta_{\text{LOS}} \Gamma (\xi)}{\sqrt{K}} \right] + {z(t)},
\end{aligned}
\end{equation}
where $\Omega = \{c_{mn}, \Delta\tau_{mn}, \tau^o_{mn}, \forall m,n\}$ represents the TM-IRS parameter configuration, and
\begin{equation}
\begin{aligned}
    V(l, \Omega, \theta, \phi) &= \sum_{m=0}^{M_x-1} \sum_{n=0}^{M_z-1} a_{mn}(\theta_b, \phi_b) c_{mn} a_{mn}(\theta, \phi) \\
    &\quad \times \Delta\tau_{mn} \mathrm{sinc}(l\pi \Delta\tau_{mn}) e^{-j l\pi(2\tau^o_{mn} + \Delta\tau_{mn})}.
\end{aligned}
\end{equation}


Here, $V(l)$ denotes the coefficient of the $l$-th harmonic generated by the time modulation of all IRS element at direction $(\theta, \phi)$. After OFDM demodulation, the received data symbol on the $i$-th subcarrier can be expressed as
%
\begin{equation}\label{eq4}
\begin{split}
    y_{i}(\theta, \phi, \xi) = &{\frac{\zeta_{\text{NLOS}} \sqrt N_t}{\sqrt K}} \sum_{k=0}^{K-1} d(k) V(i-k, \Omega, \theta, \phi) 
    \\
    &+ \frac{\zeta_{\text{LOS}} \Gamma (\xi)}{\sqrt{K}} d(i) + z_{i}.
\end{split}
\end{equation}
where  $z_{i}$ represents the overall noise contribution after demodulation. We assume that the noise is Gaussian with zero mean and variance $\sigma_u^2$ at the CU receiver.
From \eqref{eq4}, we can observe that each demodulated subcarrier symbol contains a weighted summation of symbols from all subcarriers, resulting in data scrambling across subcarriers, or say, inter-subcarrier interference.
%
In \cite{xu2025tmirs}, to ensure undistorted reception at the legitimate user, the TM parameters were selected to satisfy $V(i-k, \Omega, \theta_u, \phi_u) = 0$ for all $i \neq k$. {This  is referred to as \textit{nulling scrambling} and can be achieved via closed-form rule-based TM-IRS parameter design. 
{However, the resulting rules  do not attempt to control the magnitude of $V(0, \Omega, \theta_u, \phi_u)$ and do not consider noise; when $|V(0, \Omega, \theta_u, \phi_u)|$ is small compared to the noise level,} the signal received  by the legitimate user will still be distorted. Also, the resulting rules
cannot be readily extended to multi-user scenarios. 

\begin{remark}
During OFDM down-conversion and demodulation, the receiver filters out out-of-band harmonic replicas generated by time modulation. Only the in-band components that fold into the OFDM subcarriers contribute to the baseband expression in \eqref{eq4}, while higher-order harmonics outside the signal band do not affect the received symbols. Consequently, the scrambling stays within the original OFDM bandwidth and will not sacrifice the bandwidth.
\end{remark}

In this work, we do not aim to enforce $V_{i-k} = 0$ for all $i \neq k$ (where $V_{i-k}$ denotes $V(i-k, \Omega, \theta_u, \phi_u)$ for notational simplicity) to achieve undistorted reception. Instead, we treat $V_{i-k}$ for $i \neq k$  as interference. Let us define the signal-to-interference-plus-noise ratio (SINR) of the $u$-th legitimate user at the $i$-th subcarrier as
\begin{equation}\label{SINR}
    \mathrm{SINR}_{u,i} = \frac{\eta_{u1} |V_0|^2}{\eta_{u2} \sum_{j=i-(K-1),j \ne 0}^{i} |V_j|^2 + \sigma_u^2},
\end{equation}
where 
\begin{equation}
    \eta_{u1} = \frac{(\zeta_{\text{NLOS}} \sqrt N_t + \zeta_{\text{LOS}} \Gamma (\xi_u))^2}{K},
\end{equation}
and 
\begin{equation}
    \eta_{u2} = \frac{\zeta_{\text{NLOS}}^2 N_t}{K}.
\end{equation}
The achievable sum rate across all subcarriers can then be expressed as
\begin{equation}\label{sumC}
    C_u = \sum_{i=0}^{K-1} \log_2 (1 + \mathrm{SINR}_{u,i}).
\end{equation}
The total sum rate of $N_u$ legitimate users is
\begin{equation}\label{totalC}
    C_{\text{total}} = \sum_{u=1}^{N_u} C_u.
\end{equation}
We adopt the total achievable sum rate as the communication performance metric for the proposed TM-IRS-assisted DFRC system. Moreover, to ensure that the phase of the zero-th harmonic \( V_0 \) does not distort the received symbol constellation at legitimate directions, we need to impose a constraint on the phase of \( V_0 \):
\begin{equation}\label{phaseCons}
    |\arg(V_0(\Omega, \theta_u, \phi_u, \xi_u))| \le \rho_u \quad \forall u,
\end{equation}
where $\rho_u$ is a modulation-specific threshold. For \(\mathcal{M}\)-PSK modulation, \(\rho_u\) must be smaller than \( \pi/\mathcal{M} \). We aim to maximize the total achievable sum rate while satisfying the phase constraint for each CU to ensure reliable data recovery at legitimate directions. In contrast, for unauthorized directions—where a potential eavesdropper may reside—the achievable sum rate is not guaranteed to be optimized and the phase constraint is not guaranteed satisfied, thereby realizing directional modulation and enhancing communication security. {The above security mechanism does not depend on CSI at the transmitter side but instead leverages artificial inter-subcarrier interference induced by time modulation.}

\subsection{Radar Sensing Model}
As mentioned before, assume the approximate estimates of the target's azimuth and elevation angles relative to the IRS, denoted by $(\theta_e, \phi_e)$, and its direction relative to the BS, denoted by $\xi_e$, are available. These estimates serve as the center of a region within which the target is expected to lie and are used to optimize the radar sensing performance. 
The received signal  at potential eavesdropper from both the IRS and the BS is  give by \eqref{eq3} evaluated at $(\theta_e, \phi_e)$, and the corresponding 
radar beampattern gain during an OFDM signal period is
\begin{equation}
    \gamma_r(t) = |\zeta_{\text{NLOS}} \mathbf{a}^T(\theta_e, \phi_e) \boldsymbol{\Phi} \mathbf{U}(t) \mathbf{a}(\theta_b, \phi_b) \sqrt{N_t}+\zeta_{\text{LOS}} \Gamma (\xi_e)|^2,
\end{equation}
where $\Gamma (\xi_e) = \frac{1}{\sqrt{N_t}} \sum_{n=0}^{N_t-1} w_n e^{j n \pi \cos (\xi_e)}$.
To evaluate radar sensing performance over an entire OFDM symbol duration, we average the beampattern gain, given by
\begin{equation}
    \overline{\gamma_r} = \frac{1}{T_s} \int_0^{T_s} \gamma_r(t) dt.
\end{equation}
In practice, $\overline{\gamma_r}$ can be approximated using a finite uniform time samples as follows,
\begin{equation}\label{sensing}
    \overline{\gamma_r} \approx \frac{1}{N_s} \sum_{n=1}^{N_s} \gamma_r(t_n),
\end{equation}
where $N_s$ is the number of samples and $t_n = \frac{(n - 1) T_s}{N_s}$ is the uniform sampling instant within one OFDM symbol duration. This approximated average beampattern gain serves as the radar sensing performance metric in the TM-IRS design.

\section{Problem Formulation}\label{problem}
This section formulates the TM-IRS design problem for the DFRC system proposed in Section~\ref{system_model} to integrate sensing, communication and security simultaneously. 
In practical target tracking scenarios, the target’s location is not perfectly known at the BS and the IRS due to mobility and random fluctuations. Therefore, {we consider a setting where a coarse estimate of the target’s angle, i.e., $(\theta_e, \phi_e)$ and $\xi_e$ defined in Section~\ref{system_model}, are available at the BS/IRS, and the target/eavesdropper is assumed to be also known resided in an  angular sector $\Psi$, which is defined on grid points of the target space discretized around $(\theta_e, \phi_e)$ and $\xi_e$.} Let the set of possible eavesdropper directions be denoted as
\begin{equation}\label{eveRegion}
    \Psi = \{ (\theta_p, \phi_p), \xi_p \}, \quad p=1,2,\cdots,N_p,
\end{equation}
where $(\theta_p, \phi_p)$ and $\xi_p$ denote the $p$-th discretized spatial angle relative to the IRS and the BS, respectively, within the suspected region. $N_p$ is the total number of possible angles. To quantify security, we define the secrecy rate for the $u$-th CU as the difference between the $u$-th CU's achievable rate and the eavesdropper's rate. Let $C_e(\Omega, \theta_e, \phi_e, \xi_e)$ obtained via \eqref{sumC} denote the eavesdropper's rate at location $\{(\theta_e, \phi_e), \xi_e\} \in \Psi$. The worst-case secrecy rate for $u$-th CU is then defined based on the maximum possible eavesdropper rate over all directions in $\Psi$ as follows:
\begin{equation}\label{SecrecyRate}
    R_u(\Omega) = C_u(\Omega, \theta_u, \phi_u, \xi_u) - \max_{\{(\theta_e, \phi_e),\xi_e\} \in \Psi} C_e(\Omega, \theta_e, \phi_e,\xi_e).
\end{equation}
Our objective is to find $\Omega$ to maximize the worst-case total secrecy rate across all CUs, subject to a minimum radar sensing performance threshold $\gamma_{\text{th}}$ and the phase constraint defined in \eqref{phaseCons}:
\begin{equation}
\begin{aligned}
    \max_{\Omega} \quad & \sum_{u=1}^{N_u} R_u(\Omega) \\
    \text{s.t.} \quad & \overline{\gamma_r}(\Omega,\theta_e, \phi_e, \xi_e) \ge \gamma_{\text{th}}, \\
                      & |\arg(V_0(\Omega, \theta_u, \phi_u, \xi_u))| \le \rho_u, \quad \forall u.
\end{aligned}
\label{optFB}
\end{equation}
The above constrained optimization problem is challenging to solve due to its nonlinear, nonconvex objectives and the intractability of closed-form solutions. To address this, we propose a GFlowNet-based generative framework in the following section that efficiently samples TM-IRS configurations which maximize the desired objectives while satisfying all constraints. Unlike convex or greedy optimization methods, our approach does not rely on specific structural assumptions or relaxations, making it more flexible and broadly applicable~\cite{evmorfos2024sensor}. In contrast to supervised deep learning techniques, the proposed GFlowNets operate in an unsupervised manner and do not require large volumes of labeled data—an important advantage in DFRC scenarios, where annotated physical-layer data is often limited. Furthermore, compared with other unsupervised methods such as Markov Chain Monte Carlo (MCMC) and standard reinforcement learning (RL), GFlowNets combine the structured exploration capabilities of RL with the stability of likelihood-based training, enabling diverse and high-quality sampling with improved convergence~\cite{malkin2022trajectory}.



\section{GFlowNet-Based TM-IRS Design}\label{gflownet}


In this section, we first introduce the core principles of GFlowNets and then formulate the TM-IRS parameter design problem as a MDP to enable GFlowNets’
application. We then define a suitable reward function that incorporate both secure communication and sensing objectives under the scenario discussed previously, followed by a detailed description of the proposed GFlowNet training algorithm.

\subsection{Overview of GFlowNets}
The GFlowNet framework models the sequential decision-making process as a deterministic MDP, defined over a set of states $\mathcal{S}$, with a subset of terminal states $\mathcal{X} \subset \mathcal{S}$. An MDP satisfies the Markov property, meaning that the next state depends only on the current state and action, not on the full history of the process \cite{puterman2014markov}. At each state $s \in \mathcal{S}$, a discrete set of actions $\mathcal{A}(s)$ determines the permissible transitions, forming a directed acyclic graph (DAG) structure as shown in Fig. \ref{fig3}, where the absence of cycles ensures that the flow progresses forward without revisiting past states. A trajectory consists of a sequence of actions from the root (initial) state to a terminal state, with the possibility that different action paths may reach the same state, reflecting the non-injective structure of the graph. Rewards are only assigned to terminal states, while all intermediate states carry zero reward, i.e., $\mathcal{R}(s) = 0$ for $s \notin \mathcal{X}$. The training objective in GFlowNets is to learn a stochastic policy that induces a distribution over terminal states proportional to their associated non-negative rewards \cite{bengio2021flow}.

\begin{figure}[t]
\centerline{\includegraphics[width=2.9in]{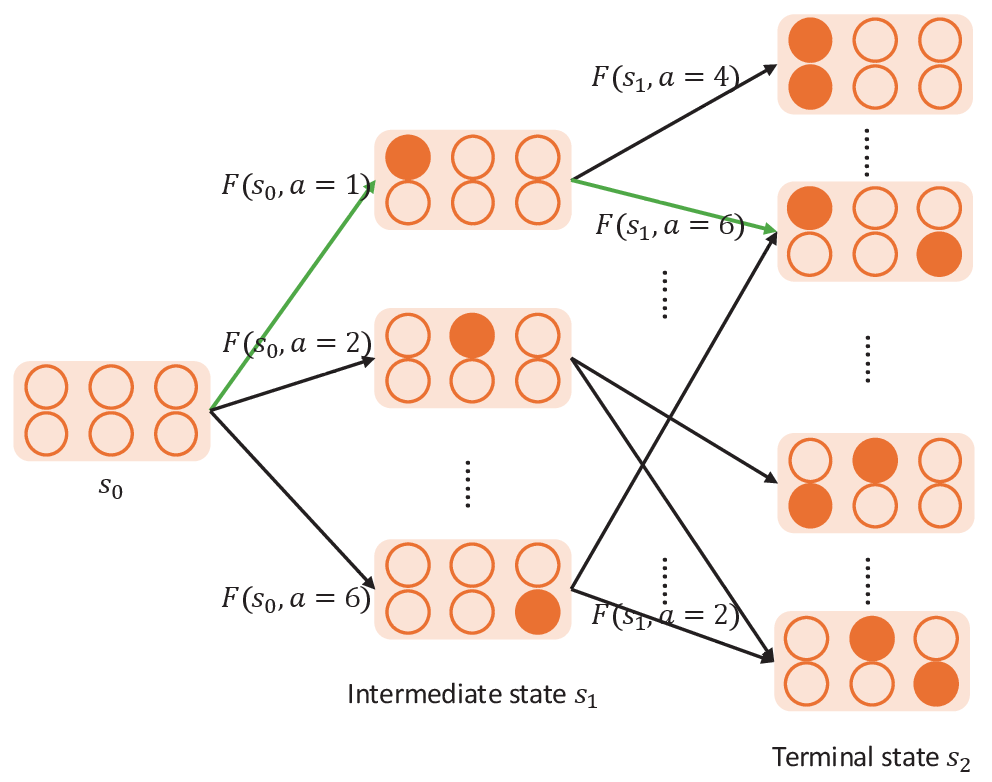}}
\caption{An example of the GFlowNet-based TM-IRS parameter selection, where two parameters are optimized, each with three discrete values. Each state represents a partially filled configuration, with solid circles indicating selected values. The green arrows highlight one trajectory from the initial to a terminal state.}
\label{fig3}
\end{figure}

\begin{figure*}[t]
    \centering
    \includegraphics[width=0.95\textwidth]{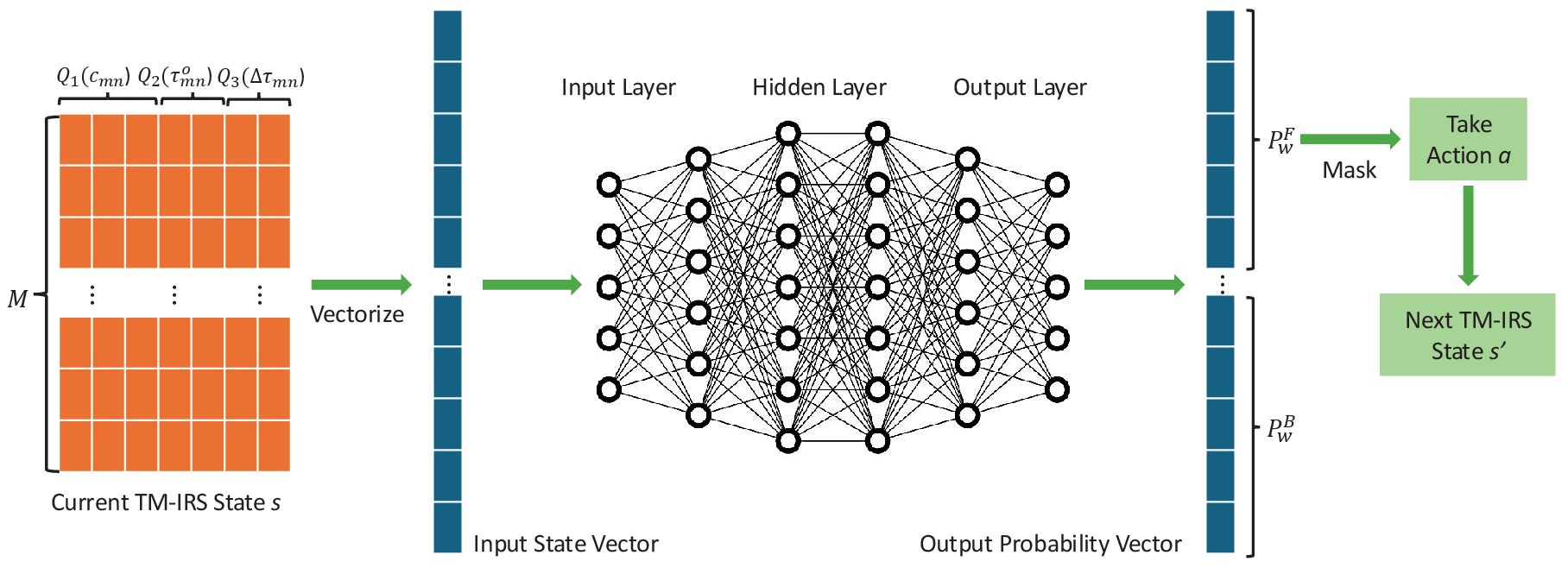}
    \caption{An illustration of the GFlowNet-based TM-IRS design framework, showing the transition from current state $s$ to next state $s'$ via deep neural network-guided action sampling.}
    \label{fig4}
\end{figure*}

To achieve this, GFlowNets view the MDP as a network of flows propagating from the root node to the terminal nodes. An edge flow $F(s, a)$ is defined for each action $a$ taken at state $s$, resulting in a transition to $s' = T(s,a)$, and the total state flow $F(s)$ corresponds to the sum of flows through that state. The flow matching principle requires that, at every state, the incoming flow equals the sum of its outgoing flow and reward. Specifically, for a node $s'$, we define the incoming and outgoing flows as
\begin{equation}
    F_{\text{in}}(s') = \sum_{s,a : T(s,a) = s'} F(s,a),
\end{equation}
\begin{equation}
    F_{\text{out}}(s') = \sum_{a' \in \mathcal{A}(s')} F(s',a').
\end{equation}
Flow conservation imposes $F_{\text{in}}(s') = \mathcal{R}(s') + F_{\text{out}}(s')$. From these flows, we define the forward and backward transition probabilities as
\begin{equation}
    P^F(s'|s) = \frac{F(s,a)}{F(s)}, \quad P^B(s|s') = \frac{F(s,a)}{F(s')},
\end{equation}
where $T(s,a) = s'$. The overall normalization constant, or partition function, of the flow network is given by the sum of rewards over all terminal states:
\begin{equation}
    Z = \sum_{x \in \mathcal{X}} \mathcal{R}(x).
\end{equation}

To train the GFlowNet, the trajectory balance (TB) loss \cite{malkin2022trajectory} is used, which considers entire trajectories from the initial to terminal states. For a sampled trajectory $\tau = (s_0 \to s_1 \to \dots \to s_n = x)$, the TB objective compares the forward and backward path probabilities, scaled by the reward and partition function:
\begin{equation}\label{TBL}
    \mathcal{L}_{\mathbf{w}}(\tau) = \left( \ln \frac{Z_{\mathbf{w}} \prod_{t=1}^n P^F_{\mathbf{w}}(s_t|s_{t-1})}{\mathcal{R}(x) \prod_{t=1}^n P^B_{\mathbf{w}}(s_{t-1}|s_t)} \right)^2,
\end{equation}
where both $P^F_{\mathbf{w}}$ and $P^B_{\mathbf{w}}$ are parametrized using deep neural networks with learnable parameters $\mathbf{w}$, and $Z_{\mathbf{w}}$ is a trainable scalar approximating the partition function. 
%
{Minimizing this TB loss enforces consistency between the forward and backward probabilities along every trajectory, scaled by the reward of the terminal state. This constraint effectively requires that terminal states with larger rewards receive proportionally larger flow through the network. As a result, the learned forward policy is pushed to generate high-reward terminal states more frequently, while low-reward states receive little flow and are sampled rarely. Consequently, the induced marginal distribution over terminal states satisfies
\[
P_{\mathrm{GFlowNet}}(x) \propto \mathcal{R}(x),
\]
meaning that the GFlowNet learns to allocate sampling probability in proportion to the reward. In short, the TB loss teaches the model to route more probability mass toward configurations with higher reward, thereby aligning the terminal-state distribution with the desired reward distribution.}

\subsection{GFlowNets for the TM-IRS Parameter Design}
We leverage the GFlowNet framework to optimize $\Omega$ for all IRS elements in our DFRC system. The TM-IRS optimization is casted first as a parameter selection problem and a discrete MDP, where each intermediate state corresponds to a partial assignment of TM-IRS parameters. Specifically, each TM-IRS parameter, including $c_{mn}$, $\tau_{mn}^o$ and $\Delta\tau_{mn}$  for each IRS element, is discretized into $Q_1$, $Q_2$ and $Q_3$ possible values, respectively, i.e., 
\begin{equation*}
\begin{split}
    &c_{mn} \in \{e^{j0}, e^{j\frac{2\pi}{Q_1}}, e^{j\frac{4\pi}{Q_1}}, \cdots, e^{j\frac{2\pi (Q_1-1)}{Q_1}} \},\\
    &\tau_{mn}^o \in \{0, \frac{1}{Q_2}, \frac{2}{Q_2}, \dots, \frac{Q_2-1}{Q_2} \},\\
    &\Delta\tau_{mn} \in \{0, \frac{1}{Q_3}, \frac{2}{Q_3}, \dots, \frac{Q_3-1}{Q_3} \}.
\end{split}
\end{equation*}
Let $M = M_x M_z$ denote the total number of IRS elements. We represent the current TM-IRS state  by a binary vector $\mathbf{s} \in \mathbb{R}^{MQ \times 1}$, which is partitioned into $M$ blocks, each having $Q = Q_1 + Q_2 + Q_3$ entries and its three sub-blocks corresponding to three TM-IRS parameters $c_{mn}$, $\tau_{mn}^o$ and $\Delta\tau_{mn}$ of one IRS element, as shown in Fig. \ref{fig4}.

Initially, at the root state, $\mathbf{s}$ is a zero  vector, meaning no any TM-IRS parameter has been assigned a value. The action space $\mathcal{A}(s)$ consists of choosing a value for one of the unassigned TM-IRS parameters. Therefore, after each action, a specific TM-IRS parameter is assigned one of its discretized values, by setting the corresponding entry in the associated sub-block of $\mathbf{s}$ to 1 while keeping all other entries in that sub-block at 0. After a sequence of $3M$ actions, a terminal state is reached where every TM-IRS parameter has been assigned exactly one value, and thus every sub-block in $\mathbf{s}$ contains a single 1. Figure 3 presents a simplified example of TM-IRS parameter selection process using GFlowNets. The reward associated with a terminal state is based on the formulated optimization objective in Section \ref{problem}, but modified to suit the GFlowNet framework. Specifically, we define the reward as
\begin{equation}\label{rewardA}
\begin{split}
    \mathcal{R} = &R_{\text{total}}(\Omega) \mathcal{H}(\overline{\gamma}_r(\Omega,\theta_e, \phi_e,\xi_e) - \gamma_{\text{th}}) \times \\
    &\prod_{u=1}^{N_u} \mathcal{H}(\rho_u - |\arg(V_0(\Omega, \theta_u^u, \phi_u^u, \xi_u))|),
\end{split}
\end{equation}
where $R_{\text{total}}(\Omega) = \sum_{u=1}^{N_u} R_u(\Omega)$, and \( \mathcal{H}(\cdot) \) is the Heaviside step function, i.e., \( \mathcal{H}(x) = 1 \) if \( x \ge 0 \), and \( 0 \) otherwise. This formulation encourages the GFlowNet to generate TM-IRS parameter configurations that maximize the legitimate communication user performance only if the phase constraint \( |\arg(V_0)_u| \le \rho_u \) is satisfied for all users and the sensing performance is guaranteed to be above the threshold. Infeasible solutions that violate any user’s phase or sensing constraint are assigned zero reward and are thus disincentivized during training. 
\begin{remark}
Although it is common in deep learning to design reward functions as additive combinations of objectives, e.g.,
\begin{equation*}
\begin{split}
    \mathcal{R} = &\lambda_c R_{\text{total}}(\Omega) + \lambda_r (\overline{\gamma}_r(\Omega,\theta_e, \phi_e,\xi_e) - \gamma_{\text{th}}) + \\
    &\sum_{u=1}^{N_u} \lambda_u (\rho_u - |\arg(V_0(\Omega, \theta_u^u, \phi_u^u, \xi_u))|),
\end{split}
\end{equation*}
where $\lambda_c, \lambda_r$ and $\lambda_u$ are the hyperparameters, we intentionally avoid such reward design for two key reasons. First, such formulations treat constraint violations as soft penalties, which do not guarantee strict satisfaction of critical requirements such as radar beampattern gain thresholds or legitimate user phase bounds. In contrast, our use of Heaviside functions enforces these constraints explicitly by assigning zero reward to infeasible configurations. Second, additive rewards introduce additional weight parameters \( \lambda_c, \lambda_r, \lambda_u \), and it usually requires substantial effort to fine-tune such hyperparameters. In contrast, our multiplicative reward structure avoids this additional tuning burden. {Moreover, the Heaviside functions in \eqref{rewardA} do not affect convergence of the GFlowNet training, since the reward is used only as a terminal scalar value for trajectory sampling and not as a differentiable objective. Hence, these Heaviside terms do not introduce non-differentiability into the TB loss, i.e., \eqref{TBL}, used for gradient descent.} 
\end{remark}

\begin{algorithm}[t]
\caption{Proposed GFlowNet-Based TM-IRS Design}
\label{alg1}
\begin{algorithmic}[1]
\State \textbf{Initialize:} Neural network parameters $\mathbf{w}$, log-partition estimate $\ln Z$, learning rate $\alpha$, batch size, initial temperature
\For{each training episode}
    \State Initialize empty state $s_0 = \mathbf{0} \in \mathbb{R}^{MQ \times 1}$
    \State Initialize trajectory buffer $\mathcal{B} = [\,]$
    \For{$t = 1$ to $T = 3M$}
        \State Compute $P_{\mathbf{w}}^F(s_t|s_{t-1})$ and $P_{\mathbf{w}}^B(s_{t-1}|s_t)$
        \State Apply temperature scaling $\frac{1}{\varepsilon}$ for $P_{\mathbf{w}}^F(s_t|s_{t-1})$
        \State Mask invalid or completed actions in $P_{\mathbf{w}}^F(s_t|s_{t-1})$
        \State Recompute $P_{\mathbf{w}}^F(s_t|s_{t-1})$ by Softmax operation
        \State Sample action $a_{t-1} \sim P_{\mathbf{w}}^F(s_t|s_{t-1})$
        \State Update state $s_t = T(s_{t-1}, a_{t-1})$
        \State Append $s_t, P_{\mathbf{w}}^F, P_{\mathbf{w}}^B$ to trajectory buffer $\mathcal{B}$
    \EndFor
    \If{$s_T$ is a valid TM-IRS configuration}
        \State Compute reward $\mathcal{R}(s_T)$ using Eq.~\eqref{rewardA}
    \EndIf
    \State Compute trajectory balance loss $\mathcal{L}(\tau)$ using \eqref{TBL}
    \State Update parameters $\mathbf{w}$ and $\ln Z$ via the Adam optimizer.
    \State Anneal temperature $\varepsilon$ based on linear decay
\EndFor
\end{algorithmic}
\end{algorithm}

The forward and backward sampling policies, $P_{\mathbf{w}}^F$ and $P_{\mathbf{w}}^B$, are modeled by a feedforward neural network parametrized by $\mathbf{w}$\footnote{While a feedforward neural network is used in this work, alternative architectures such as convolutional neural networks (CNNs) and graph neural networks (GNNs) may also be applicable and are worth investigating in future research.}, as shown in Fig. \ref{fig4}. The output of the network is a vector of dimension $2M \times Q$, where the first $M \times Q$ entries correspond to the forward transition probabilities and the latter $M \times Q$ entries correspond to the backward transition probabilities. During training, the action selection is based on the forward probabilities $P_{\mathbf{w}}^F$. To prevent repeated selection of already assigned parameters, the forward probabilities for completed parameters are masked to zero at each decision step. The network is trained using the TB loss described in \eqref{TBL}, ensuring that the learned forward policy samples TM-IRS parameter configurations with probability proportional to their associated reward in \eqref{rewardA}. To improve convergence and encourage better exploration of high-reward regions early in training, we apply a temperature annealing strategy to the logits of $P^F_{\mathbf{w}}$, scaling them by a factor $1/\varepsilon$ where the temperature $\varepsilon$ is gradually reduced over training epochs. This technique sharpens the sampling distribution over time, allowing the policy to shift from broad exploration to concentrated exploitation as learning progresses. Training is conducted offline by sampling a large amount of root-to-leaf trajectories in the MDP, applying the TB loss, and updating $\mathbf{w}$ and the total reward $Z$ via Adam gradient descent~\cite{kingma2014adam}. After training, the GFlowNet is deployed online to sample diverse high-reward TM-IRS parameter configurations. The detailed training process of GFlowNet-based TM-IRS design is exhibited in Algorithm \ref{alg1}.



\section{EXPERIMENTS}\label{experiments}
In this section, we present numerical results to illustrate both the training behavior of the proposed GFlowNet-based TM-IRS design framework and its superior performance in securing DFRC. We evaluate the effectiveness of the learned TM-IRS configurations in terms of communication reliability and secrecy performance, and compare them against several representative baselines \cite{xu2025tmirs, valliappan2013antenna, tvt2019time, li2023irs}.



\begin{figure}[t]
\centering
\subfigure[]{
\centering
\includegraphics[width=3.0in]{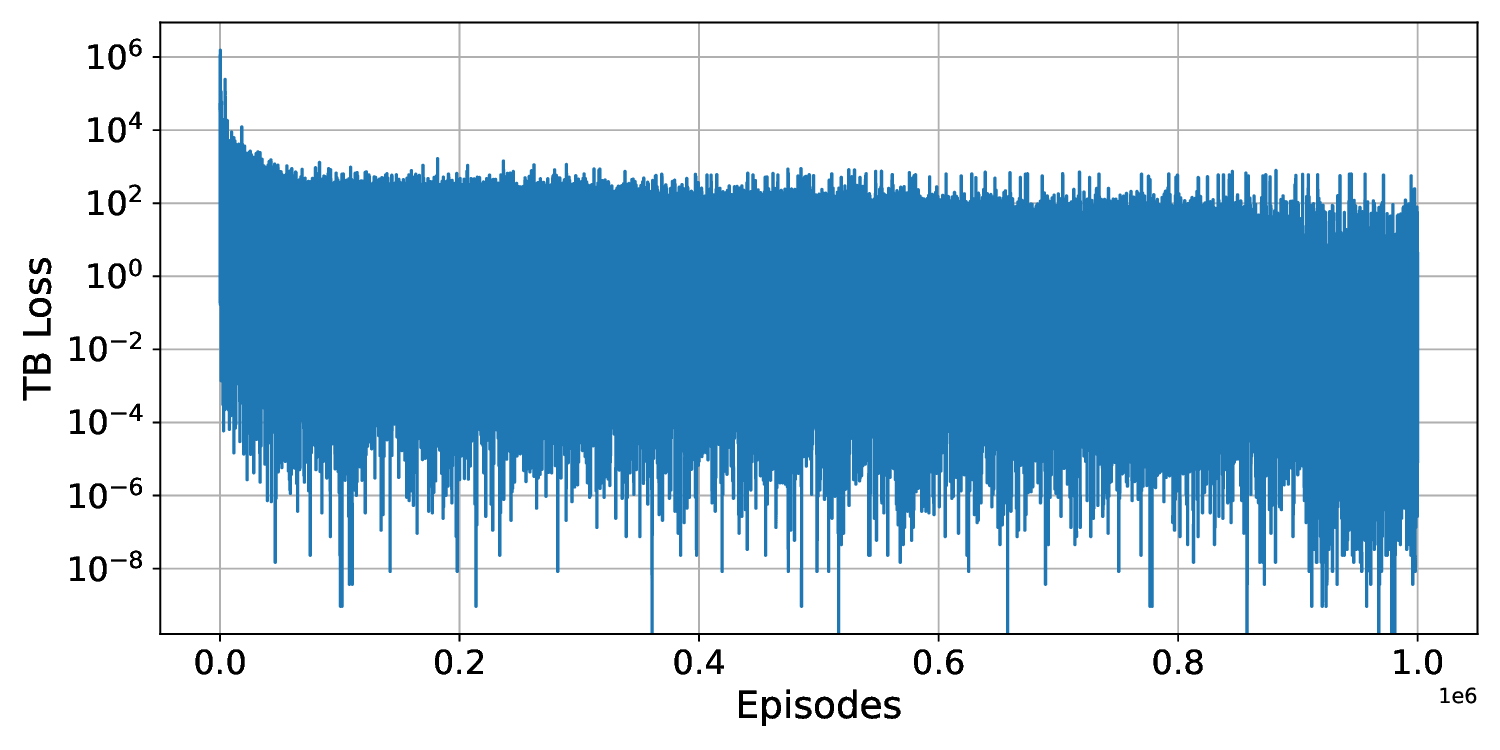}
}

\subfigure[]{
\centering
\includegraphics[width=3.0in]{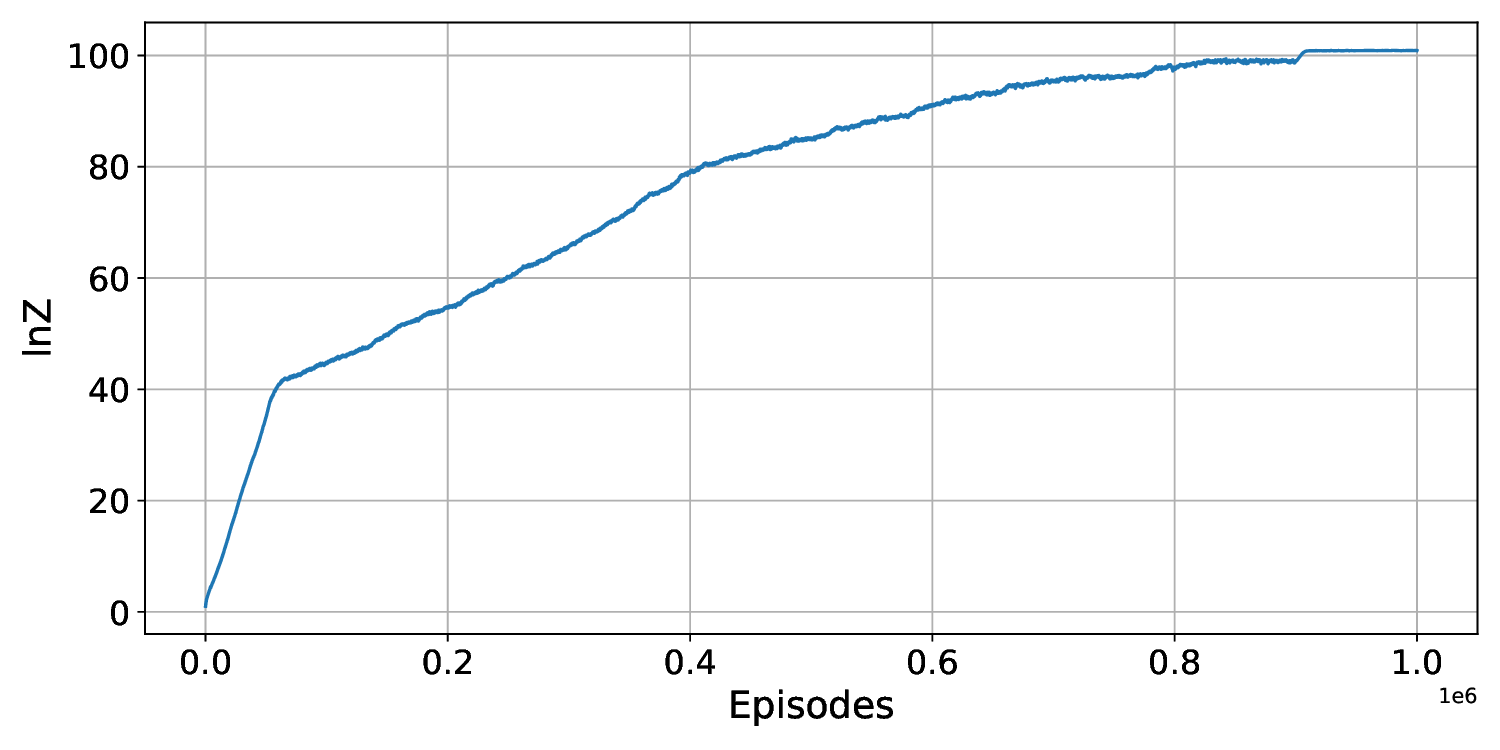}
}
\centering
\caption{Evolution of the TB loss and the estimated partition function $\ln Z$ over training episodes.}
\label{fig5}
\end{figure}

\subsection{Simulation Setup}
We consider the system of Fig.~\ref{fig1}, with
 $N_t = 8$ antennas and a square IRS with $M_x = M_z = 6$ passive reflecting elements. The system operates over $K = 16$ subcarriers, transmitting 1024 OFDM symbols and employing QPSK modulation. {The center frequency $f_c$, the subcarrier spacing $f_s$, and the OFDM duration $T_s$ are set as 24 GHz, 120 KHz, and 8.92 $\mu$s, respectively.} Unless otherwise specified, the data transmission signal-to-noise ratio (SNR) is fixed at 20 dB, $\rho_u$ is set as $ \pi/5$, and $N_p = 9$, $N_s = 8$. 
{The target (eavesdropper) and CUs are deployed at a distance of 10 m from the IRS, which represents a typical short-to-medium range deployment for indoor and urban ISAC scenarios. At $f_c = 24$~GHz (wavelength $\lambda \approx 1.25$~cm), the physical apertures of the BS ULA and the IRS are only a few centimeters. The corresponding Fraunhofer distances are approximately $0.31$~m for the BS ULA and $0.16$~m for the IRS according to the Fraunhofer distance formula. Therefore, at the 10~m distance used in the simulations, the propagation lies well within the radiative far-field region and near-field effects are negligible.
} 

To model large-scale path loss, we adopt the distance-dependent model $L(\hat{d}) = c_0 \left(\frac{\hat{d}}{d_0} \right)^{-\hat{\alpha}}$, where $c_0 = -30$ dB is the path loss at the reference distance $d_0 = 1$ m, $\hat{d}$ is the link distance, and $\hat{\alpha}$ is the path loss exponent. We set $\hat{\alpha} = 2$ for the IRS-target link, and apply Rician fading to the BS-IRS and IRS-user links with $\hat{\alpha} = 2.2$. The BS-user link follows Rician fading with $\hat{\alpha} = 2.7$.
{The initial position of the target/eavesdropper is set as \( (\theta_e, \phi_e) = (0^\circ, 0^\circ) \) with respect to the IRS, and we model the suspected region $\Psi$, defined via \eqref{eveRegion}, as a $3 \times 3$ square grid centered at $(0^\circ, 0^\circ)$, representing a localized angular area where the target/eavesdropper may reside.
} The locations of CUs will be defined on each experiment.

Each TM-IRS parameter, $c_{mn}$, $\tau^o_{mn}$ and $\Delta \tau_{mn}$, are discretized into $Q_1 = 16, Q_2 = Q_3 = 8$, respectively, unless otherwise specified. We adopt a nearest-neighbor decision method for symbol detection. A feedforward neural network with three hidden layers, each containing 256 neurons, is used to parametrize the GFlowNet, which is trained offline via an NVIDIA A100 chip with 32 GB memory and an Apple M3 Max chip with 36 GB memory.

{We use symbol error rate (SER), worst-case secrecy rate, and received signal constellation as the performance metric. We adopt SER since it provides a direct measure of constellation preservation at the CU and symbol scrambling at the eavesdropper, which aligns with the objective of the directional-modulation design.} To evaluate SER on a logarithmic scale, an offset of $10^{-4}$ is added when necessary to handle zero-SER cases. In the SER heatmaps, darker regions indicate lower error rates.

\subsection{GFlowNet Training Behavior}
We begin with a single legitimate user located at \( (\theta_u, \phi_u) = (40^\circ, 30^\circ) \) and $\xi_u = 30^{\circ}$ to demonstrate the performance of our proposed GFlowNet-based design and to facilitate a fair comparison with the rule-based TM approach in~\cite{xu2025tmirs}. 
Here \( c_{mn} \) is set as \(\left[a_{mn}(\theta_b, \phi_b)a_{mn}(\theta_u, \phi_u)\right]^{-1} \) for both of the methods, so $c_{mn}$ is not included in the GFlowNet and the training time can be reduced greatly. 
Also, the GFlowNet model is trained using $1 \times 10^6$ sampled trajectories, 
with a learning rate of $10^{-2}$ for the first $9 \times 10^5$ trajectories to accelerate the gradient descent and $10^{-3}$ for the remaining $1 \times 10^5$ to fine-tune the training.

Fig.~\ref{fig5} shows the evolution of the TB loss and the estimated partition function $\ln Z$ over training episodes. The TB loss steadily decreases, indicating that the forward and backward flows are being balanced properly. 
{The partition function $\ln Z$ (the sum of
rewards over all terminal states) gradually converges to a stable value as training progresses and  reaches convergence.} It is worth noting that the TM parameter space contains approximately $8^{72} \approx 10^{65}$ configurations, making exhaustive search infeasible. However, by parametrizing the flow using a deep neural network, the proposed framework effectively generalizes across the enormous solution space using only $1 \times 10^6$ samples (fewer than 0.000001\% of all possible configurations), inferring reward distribution even for a great deal of unvisited TM configurations.

\begin{figure}[t]
    \centering

    \begin{minipage}{0.235\textwidth}
        \centering
        \includegraphics[width=\textwidth]{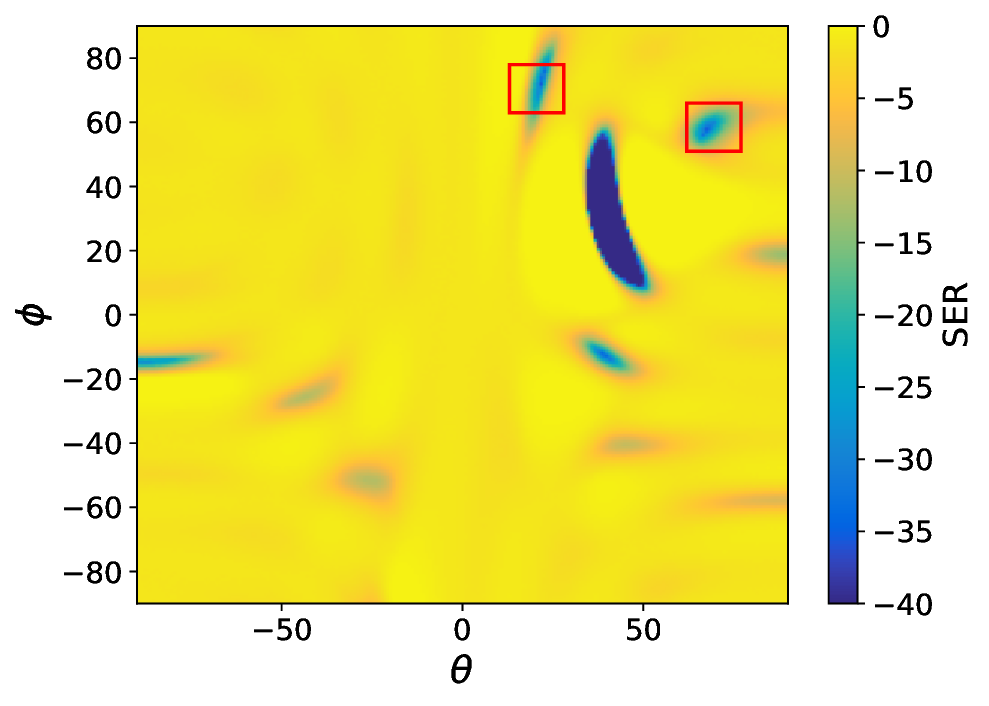}
        \vspace{0.5em}  
        \small (a)
    \end{minipage}
    \begin{minipage}{0.235\textwidth}
        \centering
        \includegraphics[width=\textwidth]{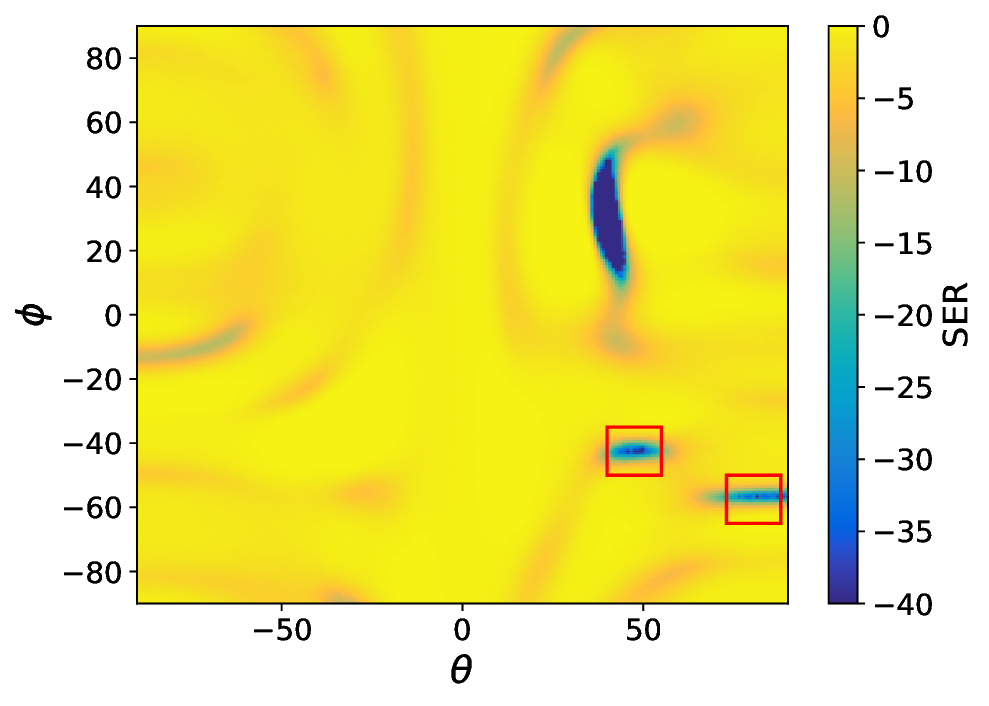}
        \vspace{0.5em}
        \small (b)
    \end{minipage}

    \caption{Comparison of SER over different spatial directions: (a) rule-based TM parameter design~\cite{xu2025tmirs}; (b) GFlowNet-based TM parameter design.}
    
    \label{fig6}
\end{figure}

\begin{figure}[t]
    \centering

    \begin{minipage}{0.235\textwidth}
        \centering
        \includegraphics[width=\textwidth]{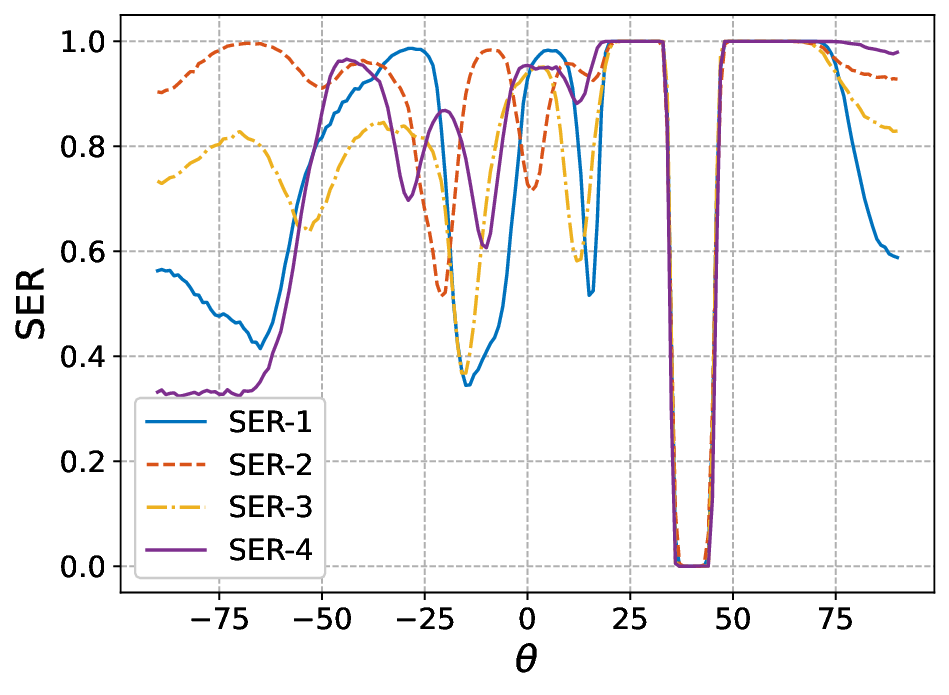}
        \vspace{0.5em}  
        \small (a)
    \end{minipage}
    \begin{minipage}{0.235\textwidth}
        \centering
        \includegraphics[width=\textwidth]{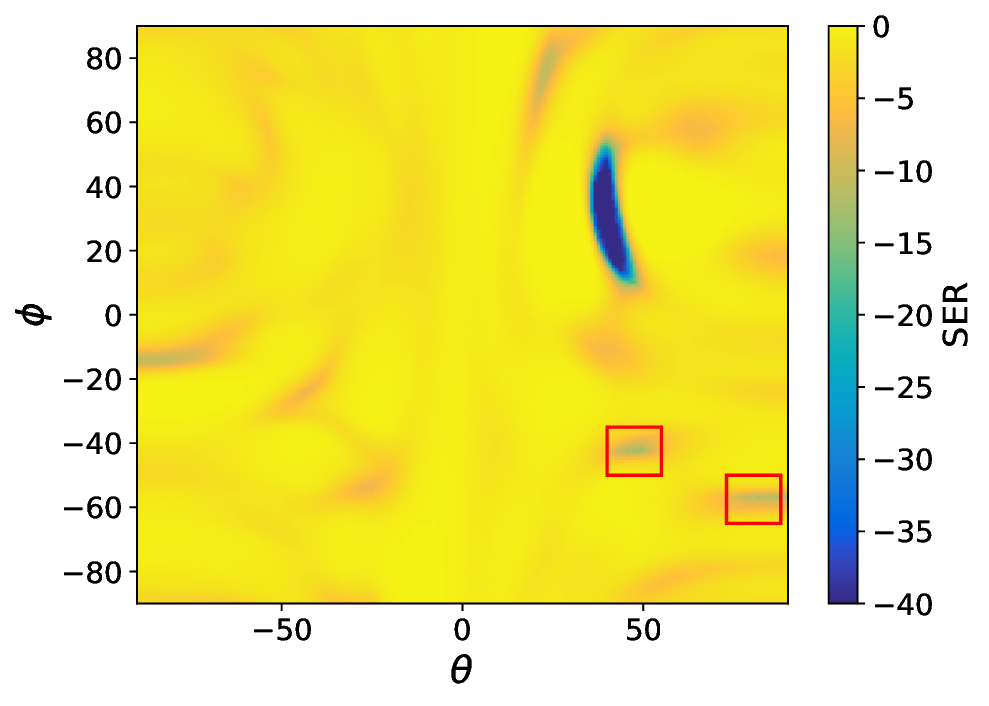}
        \vspace{0.5em}
        \small (b)
    \end{minipage}

    \caption{Enhancing security via GFlowNet diversity: (a) SER versus $\theta$ for four GFlowNet-generated TM configurations with fixed $\phi=30^\circ$; (b) averaged SER across the four configurations.}
    \label{fig7}
\end{figure}

Fig.~\ref{fig6} compares the SER performance across spatial directions for two TM design methods: the rule-based approach from~\cite{xu2025tmirs} in Fig.~\ref{fig6} (a), and the proposed GFlowNet-based method in Fig.~\ref{fig6} (b). {The angular resolution of Fig.~\ref{fig6} is $1^\circ$ for both $\theta$  and $\phi$, over the range $\theta,\phi \in [-90^\circ, 90^\circ]$.} In both cases, we can observe that the desired user direction \( (40^\circ, 30^\circ) \) achieves very low SER, while undesired directions around the target location $(0^{\circ}, 0^{\circ})$ exhibit high SER, indicating that the proposed method can achieve comparable direction modulation performance for security against the rule-based one. Moreover, several unintended directions also experience low SER, as highlighted by the red boxes in Fig.~\ref{fig6} (a) and Fig.~\ref{fig6} (b). {This arises because both  methods do not explicitly regulate the SINR in these undesired directions; as a result, certain TM-IRS configurations may inadvertently yield high SINR in those regions.} To mitigate this situation and further improve security, we can leverage the GFlowNet's capability to generate diverse high-reward TM configurations and vary the TM pattern over time. Specifically, four distinct TM parameter sets are sampled, and the configuration is switched every 256 OFDM symbols. Fig.~\ref{fig7}(a) illustrates the SER versus \( \theta \) (with fixed \( \phi = 30^\circ \)) for each of the four configurations individually. It can be seen that low-SER directions differ across configurations, while the desired user direction consistently maintains near-zero SER. Fig.~\ref{fig7}(b) shows the aggregated SER performance across all spatial directions, where the SER in previously vulnerable regions is  improved, as evidenced by the lighter color areas. This dynamic TM strategy effectively reduces the risk of eavesdropping, even when the suspicious directions are not in the vicinity of the target.

\begin{figure}[t]
\centering
\subfigure[]{
\centering
\includegraphics[width=2.8in]{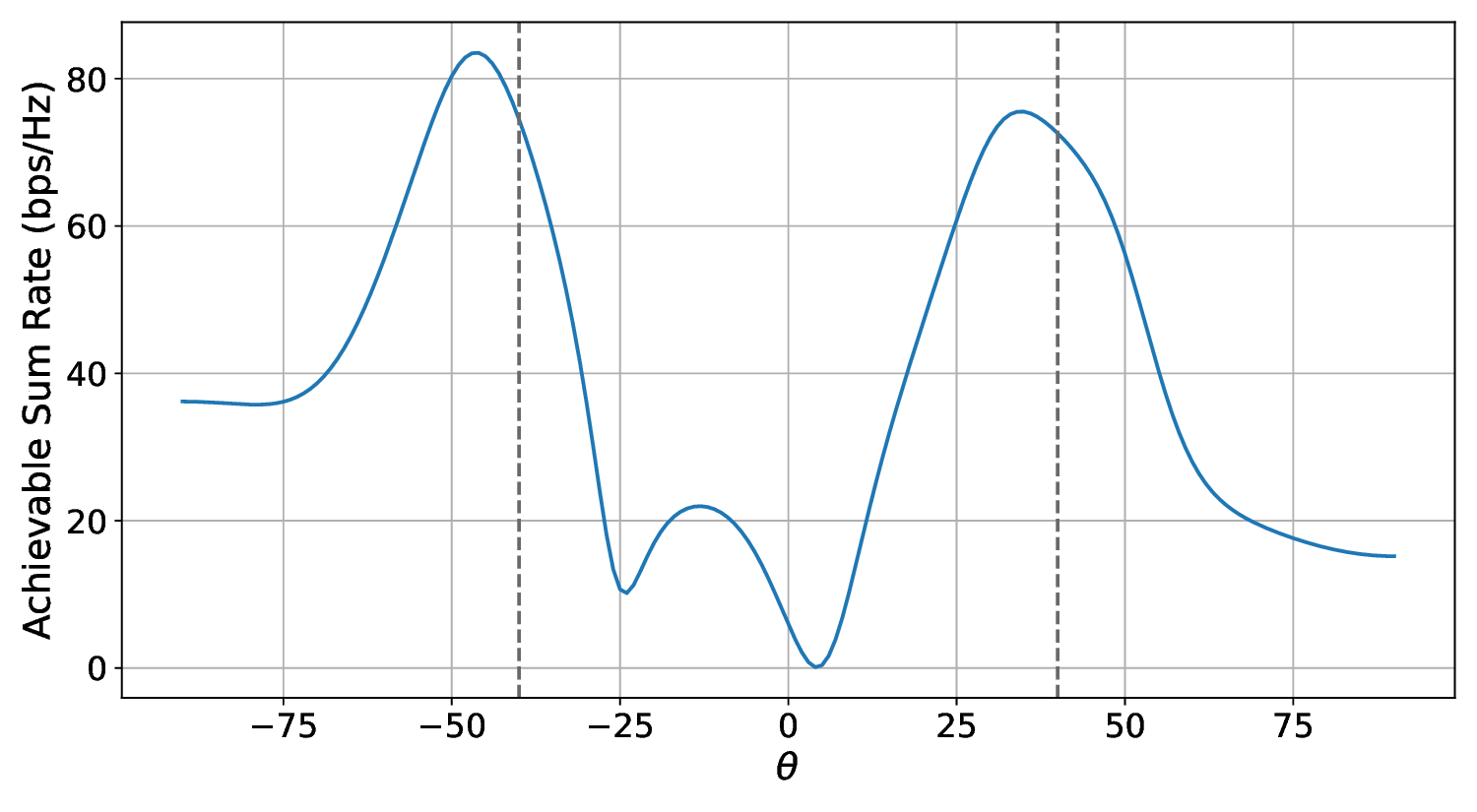}
}

\subfigure[]{
\centering
\includegraphics[width=2.8in]{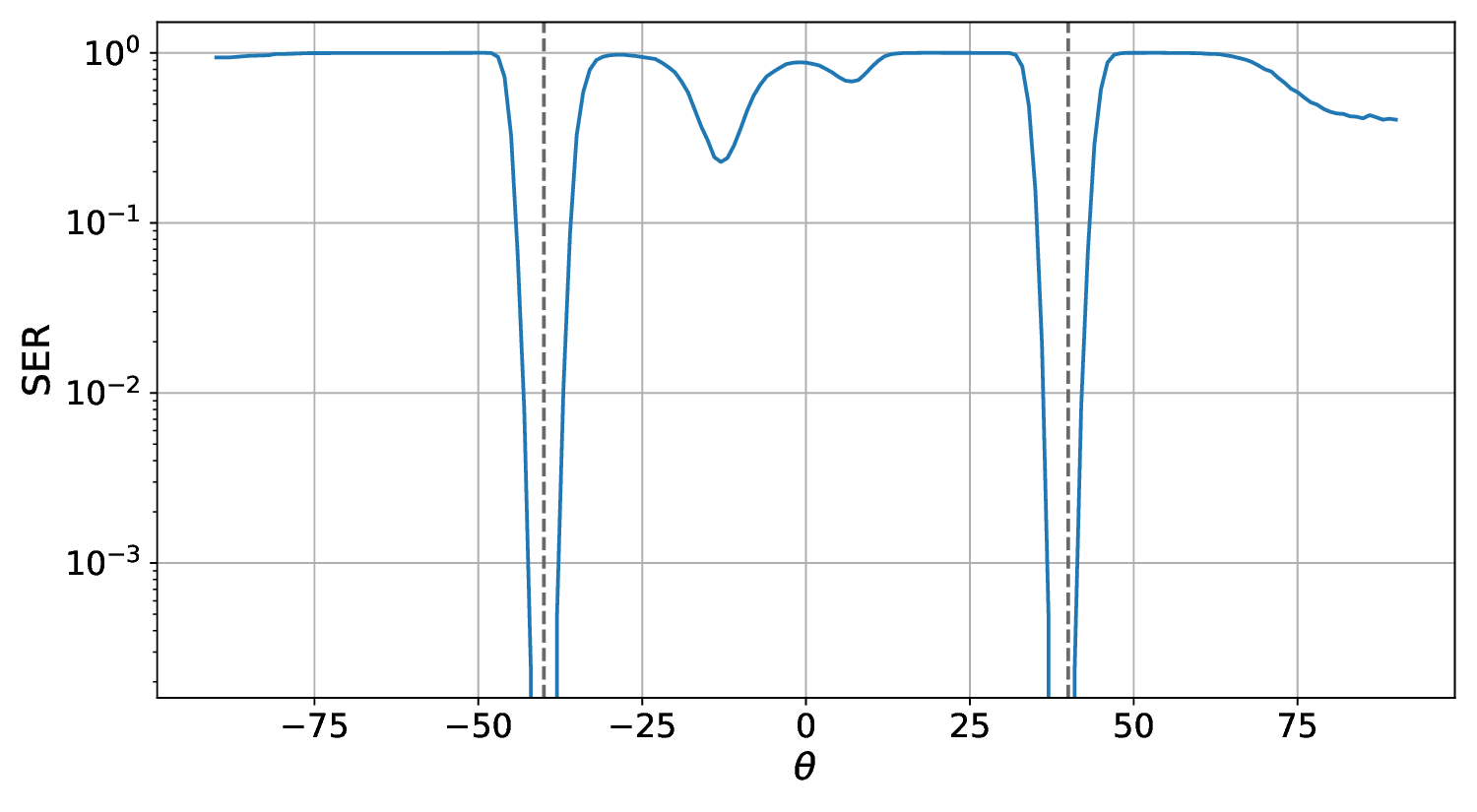}
}
\centering
\caption{A two-user scenario: (a) achievable sum rate versus $\theta$ and (b) SER versus \( \theta \) obtained via the proposed GFlowNet-based method.}
\label{fig8}
\end{figure}

\begin{figure}[t]
\centerline{\includegraphics[width=3.0in]{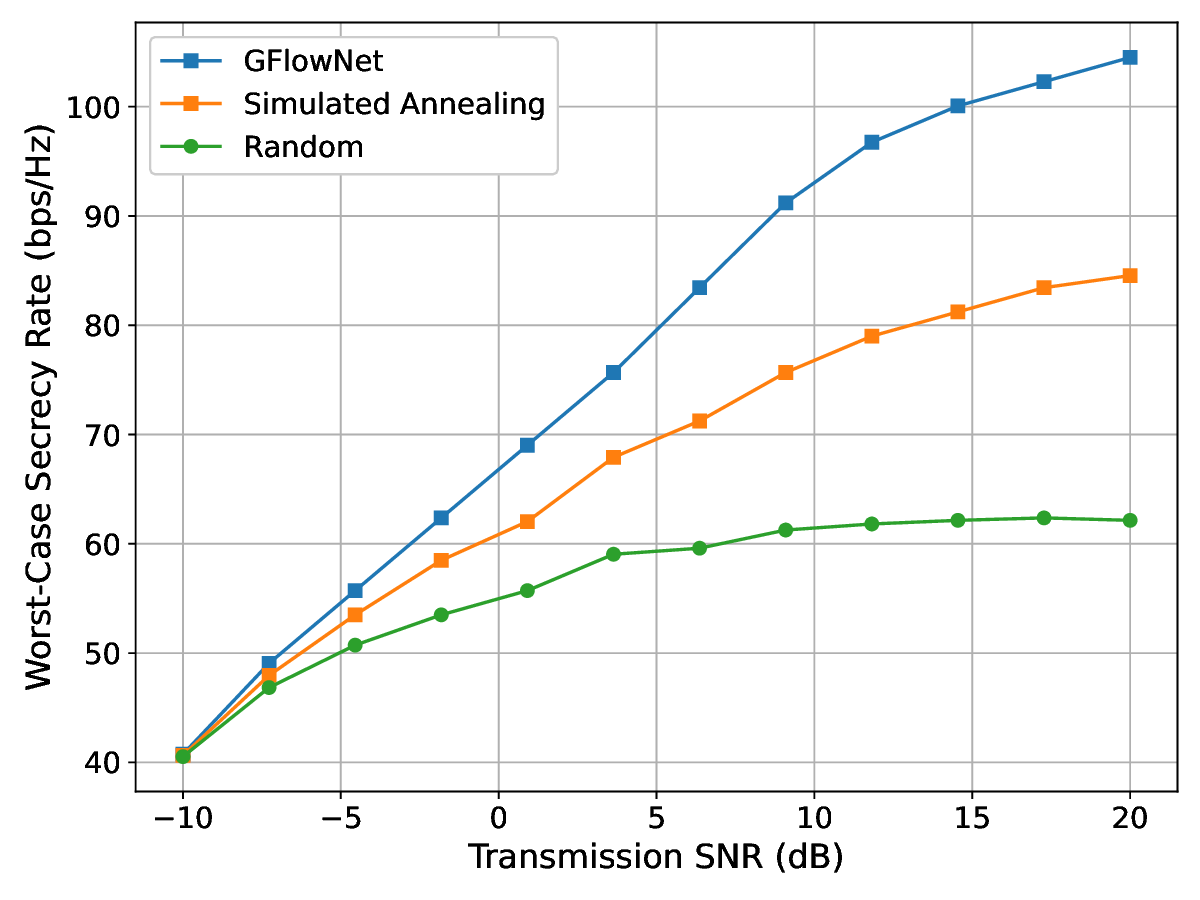}}
\caption{Comparison of worst-case secrecy rate against data transmission SNR among the proposed GFlowNet-based method and benchmarks.
}
\label{fig9}
\end{figure}

\subsection{Multi-User Security Performance}
To validate the multi-user security capability of our proposed GFlowNet-based TM-IRS design, we consider a scenario with two legitimate users located at azimuth angles \( \theta = 40^\circ \) and \( \theta = -40^\circ \), both at elevation \( \phi = 30^\circ \). Figure~\ref{fig8} illustrates the system performance across spatial angles \( \theta \), where the achievable sum rate and SER are evaluated along a 1D angular cut with fixed \( \phi = 30^\circ \) for clarity. As shown in Fig.~\ref{fig8}(a), the achievable sum rate achieves strong peaks at the desired user directions, confirming that the system supports reliable multi-user transmission. Although the obtained rates are not globally optimal\footnote{Note that the GFlowNet focuses on the probabilistic sampling instead of guaranteeing the global optimum.}, they remain high due to GFlowNet’s ability to sample TM configurations with probabilities proportional to their reward. In Fig.~\ref{fig8}(b), the SER at the two desired directions drops to near zero, demonstrating the effectiveness of the proposed method in ensuring accurate signal recovery for  multiple intended CUs while maintaining sensing performance.

{
To demonstrate that the proposed algorithm generalizes beyond specific geometric examples, we conduct experiments in which three CU locations are randomly sampled within the region $\theta \in [-40^{\circ}, 40^{\circ}], \phi \in [20^{\circ}, 60^{\circ}]$, while keeping the target region $\Psi$ unchanged. In addition, to assess the sampling efficiency of the proposed GFlowNet-based method, we compare its worst-case secrecy-rate performance across different transmission SNRs against two benchmark approaches: simulated annealing (SA), as used in \cite{valliappan2013antenna}, and uniform random search. The results are presented in Fig.~\ref{fig9}. For each SNR value, 30 sets of CU locations are drawn, and the corresponding worst-case secrecy rate is computed using \eqref{SecrecyRate}. For a fair comparison, all methods are allocated the same number of iterations. The SA algorithm treats the TM-IRS design problem as a discrete combinatorial optimization task. Starting from a random initial TM-IRS configuration, SA iteratively generates a neighboring configuration by randomly modifying one discrete TM-IRS parameter. If the new configuration yields a higher secrecy rate (i.e., higher reward), it is accepted. Otherwise, it may still be accepted with probability $\exp(\Delta \mathcal{R} / \mathcal{T})$, where $\Delta \mathcal{R}$ is the loss in reward and $\mathcal{T}$ is the temperature. This mechanism enables SA to escape poor local optima. The temperature $\mathcal{T}$ follows a standard geometric cooling schedule, starting at $\mathcal{T}_0 = 1.0$ and decaying by a factor of $0.95$ at each iteration. 
The random search method uniformly samples feasible TM–IRS configurations and greedily tracks the configuration that yields the highest secrecy-rate value among the sampled candidates. 
As shown in Fig.~\ref{fig9}, when the SNR is very low (e.g., $-10$\,dB), all schemes exhibit similarly poor performance due to noise dominance, which masks the effect of optimized TM-IRS configurations. However, as the SNR increases, a clear performance gap emerges: the proposed GFlowNet-based approach consistently achieves significantly higher secrecy rates than both SA and random search. This demonstrates its ability to efficiently explore high-reward regions of the TM-IRS parameter space, underscoring its scalability and effectiveness in high-dimensional combinatorial optimization problems.
}

{
\textbf{Complexity Analysis.} 
The random search approach  has the lowest per-iteration complexity, as it only requires uniformly sampling TM-IRS configurations and evaluating their rewards. The SA is more expensive because each step requires generating a neighboring configuration, computing the reward difference, and applying an acceptance rule under a gradually decreasing temperature schedule, leading to longer sequential optimization. By contrast, the proposed GFlowNet shifts substantial computation to an offline training stage that involves neural network updates. After training, the GFlowNet generates TM-IRS configurations via a single forward pass and can do so in parallel, resulting in significantly lower search complexity during the deployment stage compared with SA and random search.
}

\subsection{Robustness Evaluation}
To assess the robustness of the proposed GFlowNet-based TM-IRS design under challenging conditions, we conduct simulations in a low-SNR scenario with the SNR set to 0 dB and compare the SER performance against the rule-based method using only one CU, as illustrated in Fig.~\ref{fig10}. 
As observed in Fig.~\ref{fig10}, while both methods achieve lower SER at the intended direction $(\theta, \phi) = (40^\circ, 30^\circ)$ as compared to other directions, the proposed GFlowNet-based approach yields significantly lower SER values than the rule-based counterpart. This robustness can be attributed to the SINR-aware optimization adopted in the GFlowNet training process, which accounts for the magnitude of the main diagonal response $V_0$ in the SINR formulation~\eqref{SINR}. Unlike the rule-based scheme that only suppresses inter-subcarrier interference, the GFlowNet-based method simultaneously enhances the signal power and suppresses interference, yielding a stronger and more reliable signal even in low-SNR regimes. Therefore, this result demonstrates the capability of the proposed method to maintain signal quality for CU despite severe noise, which is critical in practical ISAC deployments.

\begin{figure}[t]
\centerline{\includegraphics[width=3.0in]{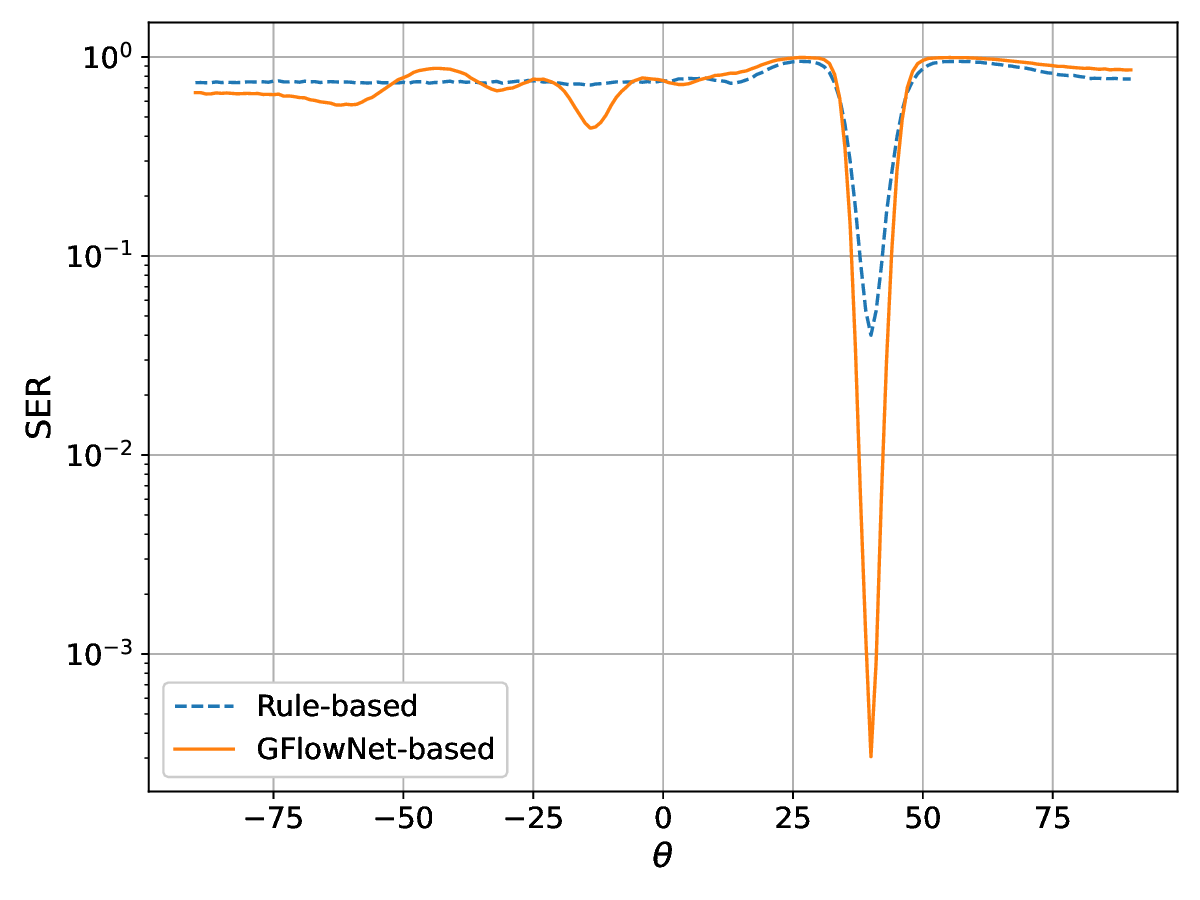}}
\caption{GFlowNet vs. the rule based TM-IRS designing in a low SNR scenario.}
\label{fig10}
\end{figure}

{
Moreover, we evaluate the resilience of the proposed method against hardware impairments, including TM switching-time jitter and IRS phase noise, both of which are unavoidable in real RF systems. In this experiment, we set the transmission SNR to 20 dB, and model hardware impairments as zero-mean Gaussian perturbations to the optimized TM switching instants and IRS phases. An impairment level of $x\%$ denotes a standard deviation of $x\% \times T_s$ for the timing jitter and $x\% \times 2\pi$ for the phase noise. Figure~\ref{fig13} illustrates the SER maps under four impairment levels: $0\%$ (the ideal case), $5\%$, $10\%$, and $30\%$, where the red star marks the CU location. As seen in Fig.~\ref{fig13}, increasing hardware impairments gradually distorts the optimized TM-IRS pattern and hence deteriorate the ideal SER map. This degradation becomes significant only at higher impairment levels (e.g., $30\%$). We can observe that the signal quality is degraded greatly around the CU and unintended low-SER regions are introduced a lot more from Fig.~\ref{fig13} (d), while from Fig.~\ref{fig13} (b), the SER patterns at $5\%$ impairment remain nearly indistinguishable from the ideal case. It is important to emphasize that the impairment levels used in the simulation are much more severe than those encountered in practical RF hardware. For example, the OFDM symbol duration in our setup is $T_s = 8.92 \,\mu\text{s}$, meaning a $5\%$ jitter corresponds to approximately $446$ ns---orders of magnitude larger than typical RF switch jitter, which is on the order of tens to hundreds of picoseconds \cite{lovelace2002effects}. Similarly, a $5\%$ phase noise level corresponds to $0.05 \times 2\pi = 18^\circ$, whereas practical phase-shifter errors are typically below $5^\circ$ \cite{kobal202128}. The results showcase the proposed method remains strong directional modulation performance even under the exaggerated $5\%$ impairment-level, indicating that our proposed design is highly robust and feasible to realistic  hardware impairments.
}

\begin{figure}[t]
\centerline{\includegraphics[width=\linewidth]{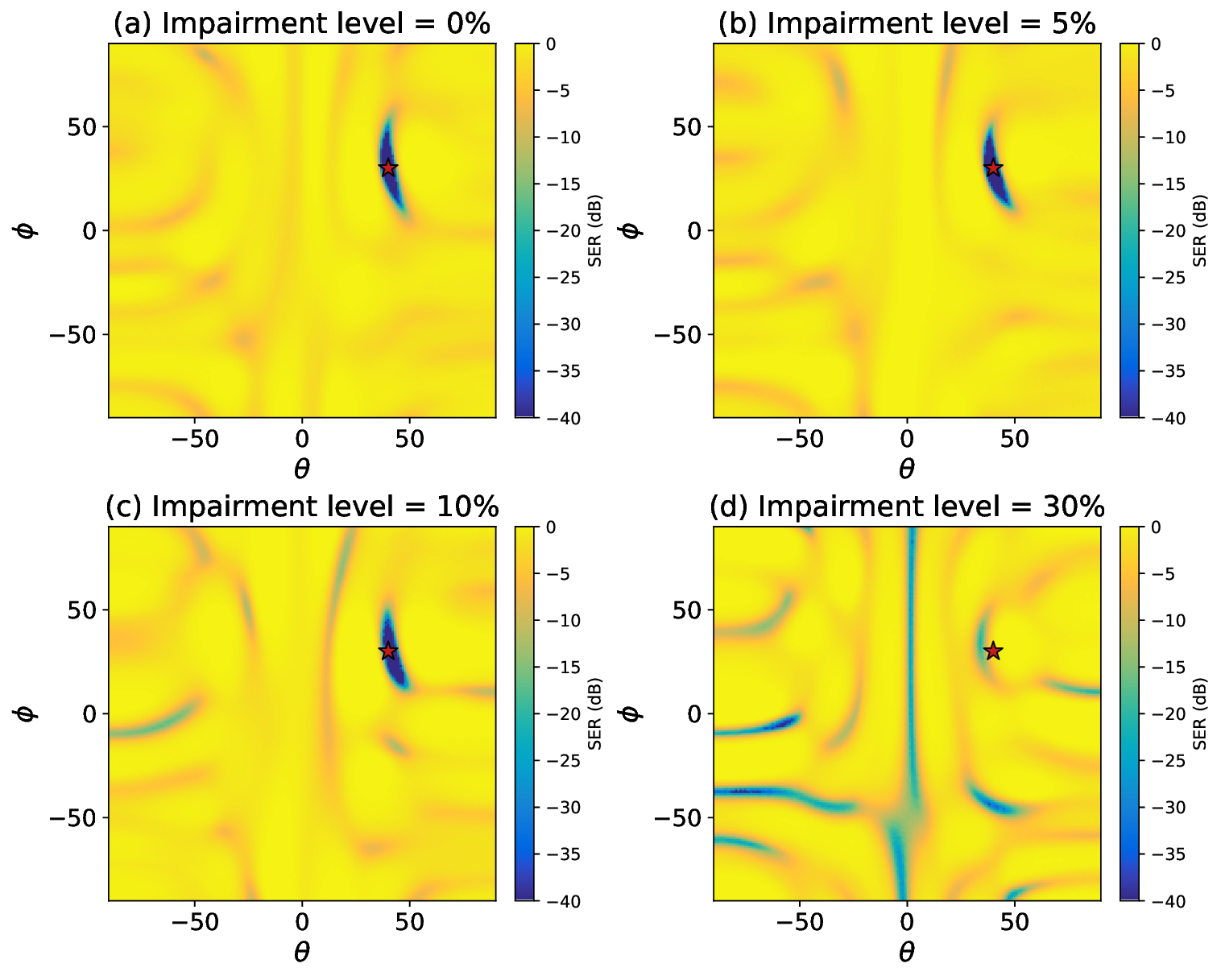}}
\caption{SER maps under different hardware impairment levels.}
\label{fig13}
\end{figure}

{
\subsection{Comparisons with Other PLS Methods}
To further demonstrate the advantages of the proposed TM-IRS--enabled security mechanism, we compare it against several representative PLS baselines. In this experiment, we consider one CU and one eavesdropper. Three benchmark schemes are examined:
\begin{itemize}
    \item \textbf{TM-ULA}: a time-modulated ULA without IRS assistance \cite{tvt2019time}, where time-modulation parameters are optimized using GFlowNets; 
    \item \textbf{Conventional ULA}: a standard ULA employing conjugate beamforming \cite{valliappan2013antenna}, which reduces sidelobe levels but does not intentionally scramble the received signal;
    \item \textbf{AN-IRS-ULA}: an IRS-aided method that employs artificial noise (AN) for PLS following \cite{li2023irs}.
\end{itemize}
Our proposed scheme is denoted as \textbf{TM-IRS-ULA}. For fairness, all methods use the same data-transmission SNR of 30\,dB, the same target/CU distance, and the same number of IRS or ULA elements. The CU is located at $(\theta_u,\phi_u) = (20^\circ,10^\circ)$ relative to the IRS and $\xi_u = 60^\circ$ relative to the BS ULA, while the eavesdropper resides at $(\theta_e,\phi_e) = (-20^\circ,-20^\circ)$ and $\xi_e = 120^\circ$.}


\begin{figure}[t]
\centerline{\includegraphics[width=3.1in]{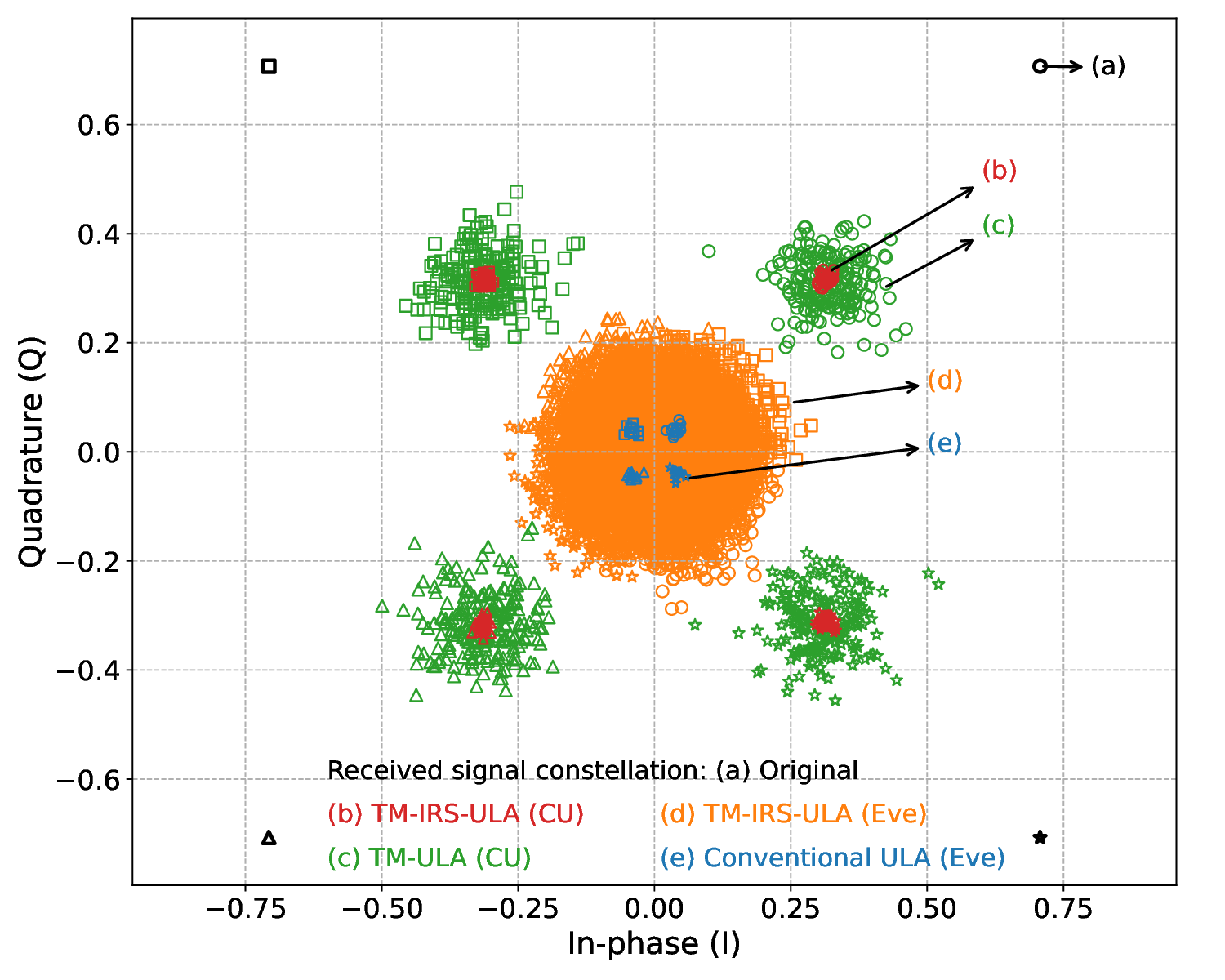}}
\caption{Received signal constellations under different PLS schemes: (a) Original QPSK constellation. (b) TM-IRS-ULA at the CU direction. (c) TM-ULA at the CU. (d) TM-IRS-ULA at the eavesdropper (Eve). (e) Conventional ULA at Eve.
}
\label{fig11}
\end{figure}

{
\subsubsection*{Constellation Performance Compared with TM-ULA and Conventional ULA}
Fig.~\ref{fig11} illustrates the received signal constellations for the CU and the eavesdropper under the TM-IRS-ULA, TM-ULA, and Conventional ULA schemes, together with the original QPSK reference constellation.}

{
At the CU direction, we observe from Fig.~\ref{fig11} that the proposed TM-IRS-ULA yields a significantly cleaner constellation than TM-ULA, even though both approaches optimize their time-modulation parameters via GFlowNets. This improvement stems from the architectural differences between the two systems: TM-ULA requires periodic deactivation of antenna elements, leading to reduced effective transmit power and degraded transmit SNR; in contrast, TM-IRS-ULA moves time modulation to the passive IRS, preserves all ULA transmit antennas, and leverages the large IRS aperture to provide stronger beamforming gain. As a result, TM-ULA exhibits notable constellation dispersion despite operating at the same 30\,dB transmit SNR, whereas TM-IRS-ULA maintains higher fidelity at the CU.
}

{
At the eavesdropper (Eve) direction from Fig.~\ref{fig11}, TM-IRS-ULA produces heavily scrambled constellations due to the inter-subcarrier interference introduced by time modulation at the IRS. Meanwhile, the Conventional ULA, which relies solely on conjugate beamforming, yields nearly undistorted symbols with only amplitude attenuation, making it vulnerable to an eavesdropper equipped with a sensitive or high-gain antenna. This contrast highlights that the time modulation-induced scrambling can provide stronger security.
}

\begin{figure}[t]
\centerline{\includegraphics[width=3.0in]{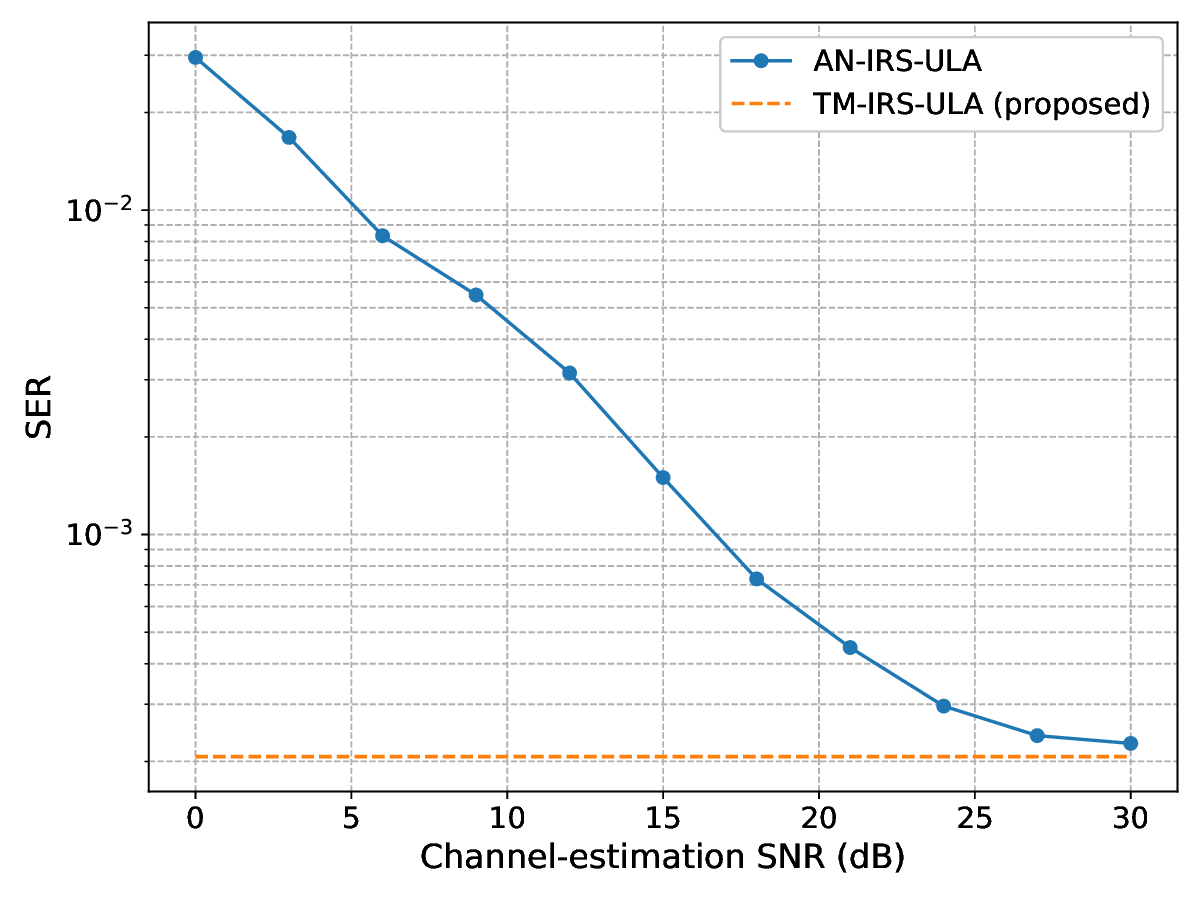}}
\caption{SER performance versus channel-estimation SNR for TM-IRS-ULA and AN-IRS-ULA.
}
\label{fig12}
\end{figure}

{
\subsubsection*{Comparison with AN-IRS-ULA Under Imperfect CSI}
We further compare TM-IRS-ULA with AN-IRS-ULA in Fig.~\ref{fig12}. The AN-based method requires accurate CSI at the transmitter to optimize the AN precoder and IRS phase shifts. To evaluate robustness, we introduce channel-estimation errors at the transmitter side and examine the CU's SER as a function of the channel-estimation SNR. Note that the receiver side is still assumed to know full CSI for decoding the data.
}

{
As shown in Fig.~\ref{fig12}, the proposed TM-IRS-ULA maintains low SER across a wide range of channel-estimation SNR values (0--30\,dB), demonstrating strong robustness even without access to CSI at the transmitter. In contrast, AN-IRS-ULA achieves low SER only when the transmitter-side CSI is highly accurate. This behavior is consistent with the SINR expression in~\eqref{SINR}: TM-IRS-ULA optimizes the OFDM subcarrier-level structure and does not rely on channel-level CSI, whereas AN-IRS-ULA computes its SINR based on channel-level information and is therefore sensitive to CSI mismatch. When the CSI is imperfect, its optimized precoders and IRS phase shifts become inaccurate, resulting in significant distortion even at the legitimate CU.
}

{
Overall, these comparisons showcase that the proposed TM-IRS-ULA achieves superior secrecy performance, higher signal quality at the CU, and greater robustness to CSI uncertainty than considered PLS baselines.
}


\section{CONCLUSION}\label{conclusion}
In this paper, we have proposed a GFlowNet-based generative framework for joint time modulation and IRS phase design in DFRC systems with secrecy constraints. Unlike previous rule-based approaches, the proposed method formulates the TM-IRS configuration task as a deterministic MDP and leverages the trajectory balance principle of GFlowNets to learn a sampling policy that generates TM-IRS parameters with probability proportional to a carefully designed reward. The reward incorporates sensing, communication and security simultaneously to satisfy our secure DFRC objective. This formulation enables unsupervised learning over a vast combinatorial space without requiring labeled data or convex approximations. To validate the effectiveness of the proposed approach, we have considered both single- and multi-user DFRC scenarios with realistic settings. Simulations demonstrate that the GFlowNet-based TM-IRS design achieves superior performance to existing rule-based and typical PLS baselines in terms of various security performance. Notably, the proposed approach provides strong security guarantees by generating diverse high-reward configurations, effectively improving security in unintended directions that are not even taken into account in the formulation. Furthermore, we have shown that the proposed method is more robust in low transmit SNR and imperfect CSI environments, highlighting its feasibility in practical ISAC systems.

Overall, this work introduces a promising GenAI framework for integrating jointly sensing, communication, and security, and opens new possibilities for learning-driven hardware designs in ISAC. Future research can extend this framework to complex ISAC scenarios by incorporating more practical factors such as hardware impairments, user mobility, etc. These challenges highlight the strength of generative AI in handling complex environments—an advantage not yet fully explored in current ISAC research. Moreover, developing more lightweight and efficient architectures for the proposed GFlowNet framework is a promising direction to reduce training overhead and enhance adaptability in future deployments.

\bibliography{npj}

\section*{Acknowledgements (not compulsory)}

This work was supported by ARO grant W911NF2320103 and NSF grant ECCS-2320568.

\section*{Author contributions statement}
Z.T. conceived the research idea, developed the methodology, conducted the simulations, and wrote the manuscript. A.P. supervised the project, contributed to the conceptual framework, provided critical insights throughout the research process, and revised the manuscript. H.V.P. contributed by reviewing the manuscript and suggesting important revisions. All authors reviewed and approved the final version of the manuscript.

\section*{Additional information}

To include, in this order: \textbf{Accession codes} (where applicable); \textbf{Competing interests} (mandatory statement). 

The corresponding author is responsible for submitting a \href{http://www.nature.com/srep/policies/index.html#competing}{competing interests statement} on behalf of all authors of the paper. This statement must be included in the submitted article file.

\end{document}